\documentclass{statsoc}
\usepackage{tabularx} % 表組みの拡張
\usepackage{subfigure} % sub-number labelling for figures
\usepackage{graphicx}
\usepackage{amsmath}
\usepackage{amsfonts}
\usepackage{enumerate}
\usepackage{url}

\usepackage{color} % added by Kato
\usepackage{bm}
\usepackage[whole]{bxcjkjatype} % added by Nakanishi
\usepackage{comment} % added by Nagasaki
\usepackage{xcite}
\usepackage{xr}

\newcommand{\kato}[1]{{\color{black}#1}} % added by Kato
\newcommand{\kator}[1]{{\color{black}#1}} % added by Kato
\newcommand{\katorr}[1]{{\color{black}#1}} % added by Kato
\newcommand{\katorrr}[1]{{\color{black}#1}} % added by Kato
\newcommand{\katos}[1]{{\color{black}#1}} % added by Kato
\newcommand{\katoss}[1]{{\color{black}#1}} % added by Kato
\newcommand{\katosss}[1]{{\color{black}#1}} % added by Kato
\newcommand{\katot}[1]{{\color{black}#1}} % added by Kato
\newcommand{\katott}[1]{{\color{black}#1}} % added by Kato
\newcommand{\katou}[1]{{\color{black}#1}} % added by Kato
\newcommand{\katov}[1]{{\color{black}#1}} % added by Kato
\newtheorem{algo}{Algorithm}[section]
\newcommand{\argmin}{\mathop{\rm argmin}\limits}
\newtheorem{prop}{Proposition}
\newtheorem{theorem}{Theorem}

\newcommand{\cyan}[1]{\textcolor{black}{#1}} % added by Nakanishi
\newcommand{\kota}[1]{\textcolor{black}{#1}} % added by Nagasaki
\newcommand{\kotan}[1]{\textcolor{black}{#1}} % added by Nagasaki
\newcommand{\kotann}[1]{\textcolor{black}{#1}} % added by Nagasaki
\newcommand{\kotar}[1]{\textcolor{black}{#1}} % added by Nagasaki
\newcommand{\kotarr}[1]{\textcolor{black}{#1}} % added by Nagasaki
\newcommand{\chris}[1]{\textcolor{black}{#1}}
\newcommand{\chriss}[1]{\textcolor{black}{#1}}

\title[Traffic Count Data Analysis Using Mixtures of Kato--Jones Distributions]{Traffic Count Data Analysis Using Mixtures of Kato--Jones Distributions}
\author[Kota Nagasaki]{Kota Nagasaki}
\address{Department of Civil and Environmental Engineering, Tokyo Institute of Technology, Tokyo, Japan}
\email{nagasaki.k.ab@m.titech.ac.jp}

\author[\kato{Shogo Kato}]{\kato{Shogo Kato}}
\address{\kato{Institute of Statistical Mathematics, Tokyo, Japan}}
\email{\kato{skato@ism.ac.jp}}

\author[Wataru Nakanishi]{Wataru Nakanishi}
\address{School of Geosciences and Civil Engineering, Kanazawa University, Ishikawa, Japan}
\email{nakanishi@se.kanazawa-u.ac.jp}

\author[\chris{M.C. Jones}]{\chris{M.C. Jones}}
\address{\chris{School of Mathematics and Statistics, The Open University, Milton Keynes, U.K.}}
\email{\chris{m.c.jones@open.ac.uk}}

\begin{document}

\begin{abstract}
\katorr{
\katorrr{We discuss the modelling of traffic count data that show the variation of \katoss{traffic} volume within a day.
\katoss{For the modelling,} we apply mixtures of Kato--Jones distributions in which each component is unimodal and affords a wide range of skewness and kurtosis.
We consider two methods for parameter estimation, namely, a modified method of moments and the maximum likelihood method. 
These methods \katoss{were} seen to be useful for fitting the proposed mixtures to \katoss{our data.}}
\kota{As a result, \katoss{the variation in traffic volume was} classified into the morning and evening traffic \chris{whose distributions have different shapes, particularly} different \katoss{degrees of} skewness and kurtosis.}}
\end{abstract}


\section{Introduction}
Traffic flow volume is one of the most essential variables in \chris{the} transport engineering field.
Also, it is easy to observe by installing \chris{a} traffic counter and counting the number of vehicles \chris{passing} by at the observation site.
Many traffic counters exist along arterial roads in Japan and other countries, and the traffic flow data are accumulated day by day \citep{led08,ana13}.

\chris{These} data are mainly utilized in the following two manners.
The first is evaluating road performance, or throughput, i.e., the maximum number of vehicles the road can accommodate within a unit time \citep{edi63,hcm}.
For example, this value is represented in \chriss{units} of vehicles per hour.
Traffic congestion occurs if more vehicles than this value come to the road.
The second is the so-called ``traffic state estimation'', which \chris{comprises} three essential variables: the traffic flow rate, the average vehicle density, and the average vehicle speed.
Understanding the relationship among these three variables is quite important to predict traffic congestion that \chris{is} spread along \chris{the} road network \citep{mun03,wan05}.
Both of \chris{road performance and traffic state estimation} deal with the observed traffic flow volume as aggregated data such as 1- or 5-\chris{minute} flow \chris{volumes}, i.e., the number of vehicles \chris{that} pass by in such time intervals.
In addition, some studies use such data without aggregation, i.e., the raw timestamp of each vehicle's passing.
They also aim at analyzing congestion and modeling the time intervals \chris{between} successive vehicles \citep{che10,li17}.

Utilizing the data in such ways is reasonable regarding the purpose of installing the traffic counter.
Nonetheless, the raw data may have much potential to bring additional insights to transport engineering.
For example, if we model the probability distribution of \kotar{time of} \chris{vehicles} passing, we can interpolate missing values, which may be caused by the failure of counters.
\kotar{Predicting} the traffic volume \kotar{in the whole road section including} unobserved locations may also be possible \kotar{by combining our model with interpolation methods}.
Additionally, it is useful for the verification and validation of microscopic traffic simulations that deal \chris{with} each vehicle as an agent.
To the best of \chris{the} authors' knowledge, no study except one written in Korean \citep{na11} estimates the probability distribution of traffic volume.

In this study, considering such potential applications, we estimate the \chris{complex} distribution of the variation of traffic volume within an average weekday.
Traffic volume is generally determined by the dynamic system of the interaction between the supply side (\kotar{i.e., how many vehicles the road can accommodate}) and the demand side (i.e., how many drivers wish to use the road).
Due to this complexity, the distribution of traffic volume in a day is not expected to be unimodal or symmetric.
\kotann{Generally, the timestamp of each vehicle's passing does not follow \chris{a} Poisson process because of some bottlenecks such as traffic congestion and traffic \chris{lights} \citep{dag97}.}
Actually, it is empirically bimodal and asymmetric on weekdays (explained in detail in Section \ref{data}).
\kota{In addition, if we consider an average weekday, time is periodic and represented as points on \chris{a} circle.
Thus, the time axis is circular in that 0 a.m. and 12 p.m. represent the same time.}
This suggests \chris{that} we had better employ directional statistics.

Our study is the first attempt to use the raw data of \chris{a} traffic counter and employ \chris{a} mixture of \katot{the circular distributions of \citet{kat15}} to represent the distribution of traffic volume in an average weekday.
Also, we contribute \chris{a} method of parameter estimation for \chris{mixtures} of Kato--Jones \chris{distributions} as there is no established method thus far.
\katorr{The use of mixtures of \katot{the distributions of \citet{kat15}} has been briefly discussed for an analysis of another traffic \chris{dataset} in the conference proceedings of \citet{nag19}.
However, despite its simplicity, it is not \chris{the case} that their inferential algorithm for the mixtures guarantees the consistency or any other optimal property of the proposed estimator.}

\katot{Some mixtures of circular distributions have been proposed in the literature.}
The most attention \chris{has} been paid to mixtures of von Mises distributions \citep[e.g.,][]{wal00,moo03,ban05,mul20}.
The components of \chris{these} mixtures, the von Mises distributions, are symmetric distributions with two parameters controlling location and mean resultant length.
Recently, \citet{miy20} discussed mixtures of sine-skewed distributions whose components can adopt mildly asymmetric shapes.
However, none of these existing models seem to be appropriate for our traffic data because one of the clusters \chris{in} our data is strongly skewed.
Therefore, for the modelling of our data, it seems reasonable to consider the distributions of \citet{kat15} as possible components of the mixtures because \chris{these distributions} can control a wide range of skewness as well as kurtosis.

The paper is organized as follows.
\chris{The traffic} count data of interest are introduced in Section \ref{data}.
We define the mixtures of the distributions of \citet{kat15}, which are applied to the traffic data, and investigate basic properties of the mixtures and their components in Section \ref{sec:model}.
Two methods for parameter estimation for the proposed mixtures are presented in Section \ref{sec:estimation}.
A simulation study is conducted to compare the two proposed methods for parameter estimation in Section \ref{sec:simulation}.
The proposed mixtures are applied to the traffic data using the proposed inferential methods in Section \ref{sec:application}.
The interpretation of the estimated model is also discussed.
The fit of the proposed mixtures to the data is compared with the fits of some other mixture models in Section \ref{sec:comparison}.
Finally, Section \ref{sec:conclusion} concludes the paper.
The online Supplementary Materials provide the additional data analysis, a simulation study, proofs of the claims made in the main article, and Python codes for parameter estimation.
%proofs of Theorem 1 and Proposition 1 and Python codes for parameter estimation and simulation study.

\section{The Data}
\label{data}
First, we briefly explain about traffic counters in general.
Traffic counters record the timestamps of all \chris{vehicles} passing. 
Many counters are installed along the arterial roads in Japan, \chris{often} at intervals of one kilometer.
Usually, these data are aggregated to 1- or 5-\chris{minute} vehicle counts.
Then, the aggregated data are utilized for traffic control such as detecting congestion and providing information to drivers.
On the other hand, we use the data without aggregation in this paper, the data \chris{being} raw timestamps, to model the probability of \chris{a} vehicle passing at \chris{each particular time}.
In particular, this paper focuses on the averaged \chris{distribution for} weekdays.
Then, the time axis is circular; 0:00 (0 a.m.) and 24:00 (12 p.m.) represent the same time, and 1:00 (1 a.m.) and 23:00 (11 p.m.) are not 22 hours apart, but only 2 hours apart.
To consider \chris{these} characteristics accurately, we employ the concept of directional statistics.

\begin{figure}
\centering
\includegraphics[clip,keepaspectratio,width=0.7\textwidth]{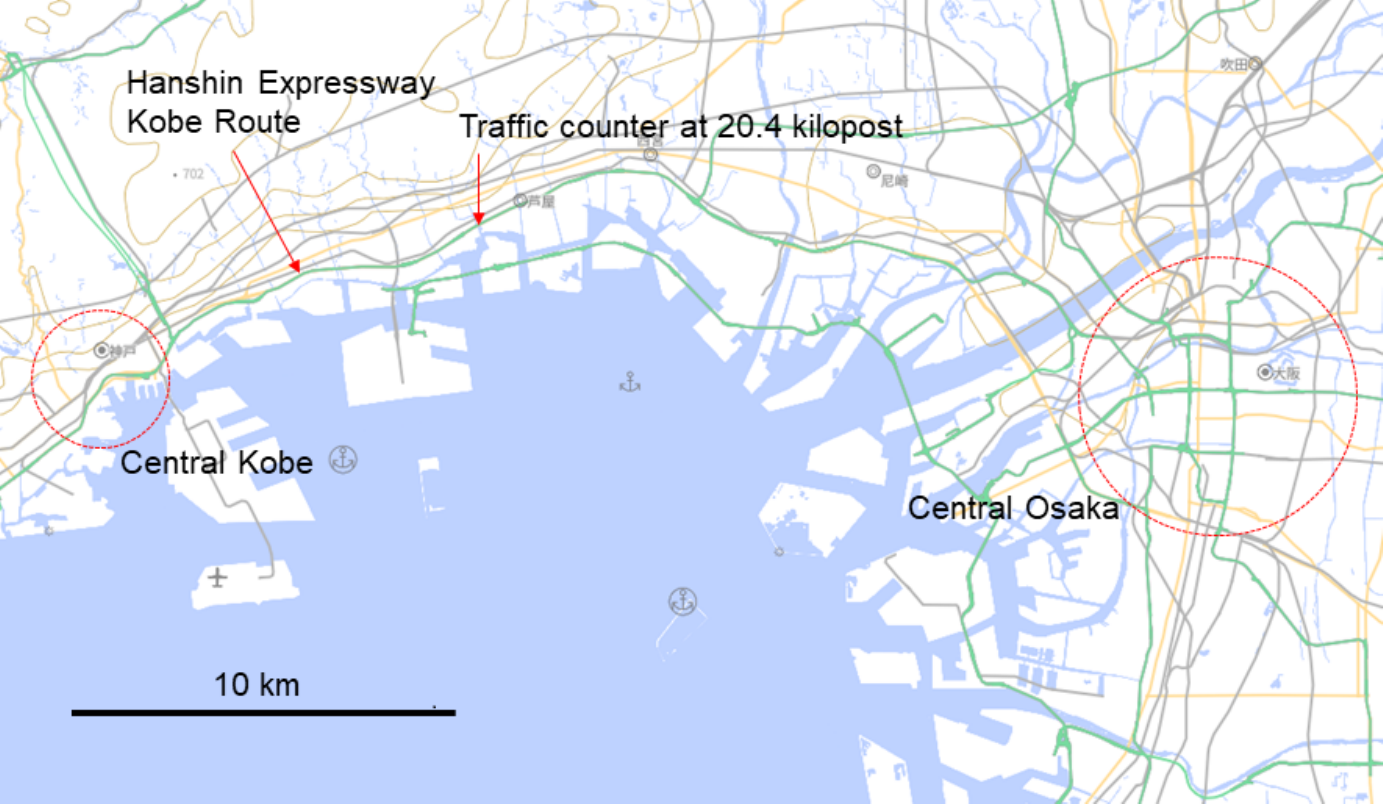}
\caption{\label{fig:target area} Map of target area \katot{including the location of the traffic counter and \chris{the} Kobe route of \chris{the} Hanshin Expressway}.}
\end{figure}

\chris{Concretely}, we use the data of a traffic counter provided by \chris{the} Hanshin Expressway Co. Ltd., Japan \chris{via} personal communication.
Specifically, the data are obtained at \chris{the} 20.4 kilopost of \chris{the} Kobe route, Hanshin \katot{Expressway.
Fig.\ \ref{fig:target area} shows \chris{a} map of the target area including the location of the traffic counter and \chris{the} Kobe route of \chris{the} Hanshin Expressway.
\chris{The} background map is based on \chris{a} Digital Map by \chris{the} Geospatial Information Authority of Japan.}
\katot{\chris{The} Kobe route} exists in \chris{the} Osaka metropolitan area and connects two large cities, Osaka and Kobe.
\chris{The} Osaka metropolitan area is the second largest megacity in Japan following \chris{the} Tokyo metropolis and its population is around 20 \chris{million}. 
Osaka city is the \chris{main} city of \chris{the} Osaka metropolitan area with \chris{a} population of around 2.8 \chris{million}.
Kobe city is the second largest city in the area with \chris{a} population of 1.5 \chris{million} and \chris{is} located about 30 km west of Osaka city.

\begin{figure}
\centering
\includegraphics[clip,keepaspectratio,width=0.75\textwidth]{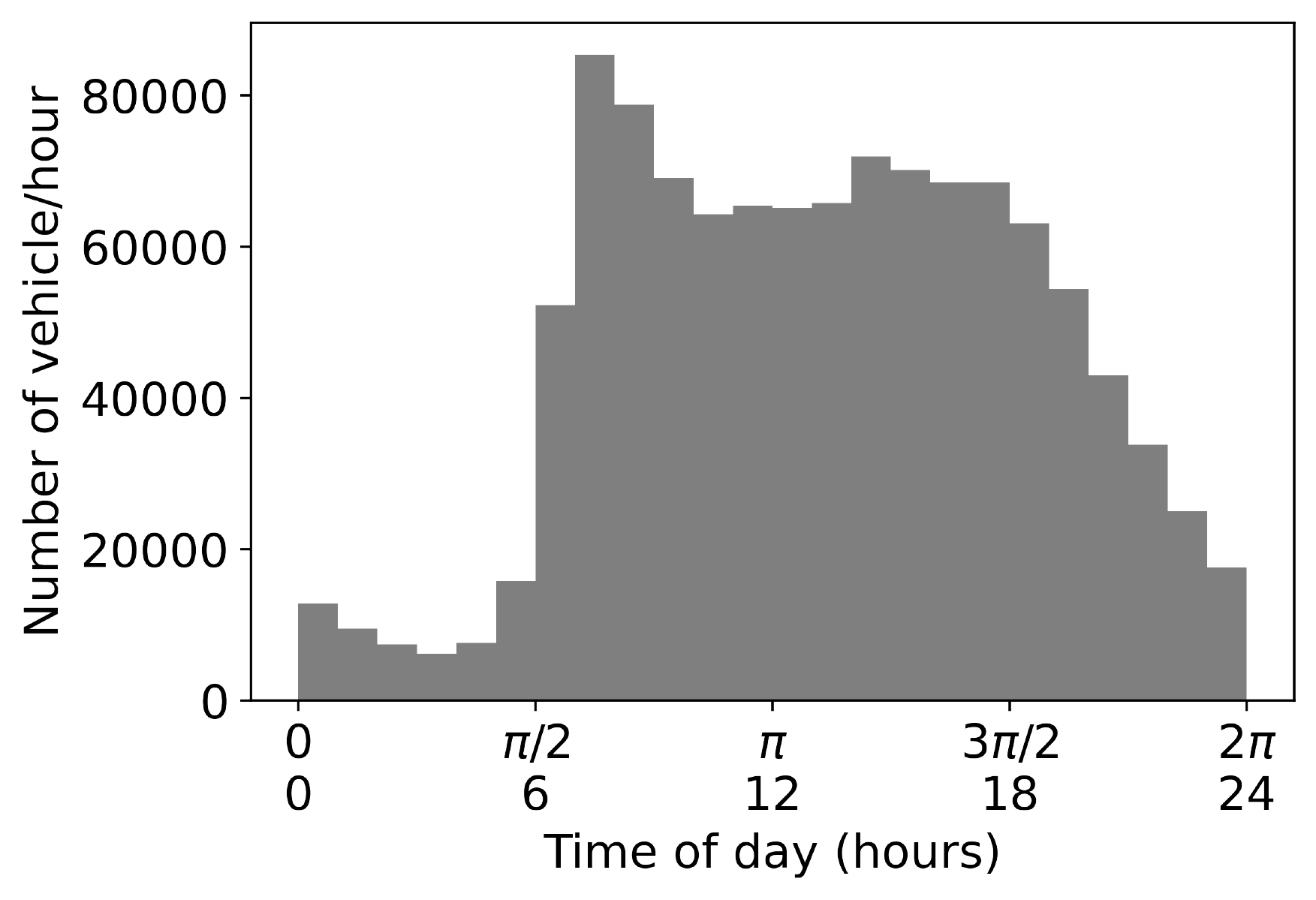}
\caption{\label{fig:weekday} \chris{A} histogram of the data for the summation of all 46 weekdays.}
\end{figure}

The counter records the timestamps of \chris{vehicles} passing in the direction from Kobe to Osaka.
The data period is 46 weekdays; from June 6th to July 7th and from August 22nd to October 10th in 2016.
\kotan{\katot{It is not unnatural to aggregate data for all weekdays because the \chris{distribution} of weekday traffic is generally similar \citep{hcm}.}}
\chris{A} histogram of the data of the summation of all 46 days \chris{is} shown in Fig.\ \ref{fig:weekday}. 
The bin width is set as 1 hour for this figure.

The total number of observed vehicles for 46 days is 1,121,262, which equals 24,375 per day on average.
Here, the histogram has two peaks at around 7:00 and 15:00.
We assume that the former peak represents the morning rush hour and the latter does the evening rush hour, from Kobe to Osaka, as the data are for weekdays.
\kotar{Note that all recorded traffic is in the same direction (i.e., from Kobe to Osaka).
It means that vehicles commuting to their office and returning home are observed only once in the morning or evening.}

In addition, the peak in the morning is higher and sharper than that in the evening.
Also, each peak has \chris{an} asymmetric shape.
For example, as many commuters should arrive \chris{at} their \chris{offices} in Osaka at around the same \katot{time, typically, at 8:30 or 9:00,} the histogram of the morning peak suggests \chris{a} negatively-skewed shape.
\kotar{A model that cannot represent flexible shapes cannot adequately describe traffic data's shape and characteristics.}
To fully represent the characteristics of this distribution, a mixture of \katot{probability density functions} that allows a flexible shape is necessary.
\kotar{
Hereafter, we assume that the data follow the IID property.
Although this might be a strong assumption, it seems acceptable from the transportation engineering viewpoint.
See Appendix A of Supplementary Material for the details.
}

\section{Model} \label{sec:model}

\subsection{Definition of the model}
This section considers a model for the traffic data.
As displayed in \katot{Fig.\ \ref{fig:weekday}}, our data show two distinct features, namely, (i) \chris{bimodality} and (ii) different degrees of skewness and peakedness in the clusters of the morning and evening rush hours.
Therefore it seems reasonable to adopt a \katot{mixture model} whose components have flexibility in terms of skewness and peakedness.
\katott{In this paper we adopt the distribution of \citet{kat15} as each component of the mixture.
Its density is given by
\begin{equation}
	g_{\rm KJ}(\theta; \mu, \gamma , \lambda, \rho )  = \frac{1}{2\pi} \left\{ 1 + 2 \gamma \, \frac{\cos 
		(\theta-\mu)  - \rho \cos \lambda}{1+\rho^2-2 \rho \cos 
		(\theta-\mu-\lambda)} \right\},\quad \katou{ 0 \leq \theta < 2 \pi,}
	\label{eq:kj_density}
\end{equation}
where the ranges of the parameters are $ \katou{0  \leq \mu < 2 \pi,} \ 0 \leq \gamma <1, \ 0 
\leq \rho <1$, and \katou{$0 \leq \lambda < 2\pi$} satisfying 
$(\rho  \cos \lambda - \gamma)^2 + (\rho \sin \lambda)^2 \leq (1 - 
\gamma)^2.$
Throughout the paper the distribution (\ref{eq:kj_density}) will be called the Kato--Jones distribution.

Then we propose the following mixture with density
\begin{align}
	f(\theta) =  \frac{1}{2\pi} \sum_{k=1}^m \pi_k \, g_{\rm KJ}(\theta; \mu_k, \gamma_k , \lambda_k, \rho_k ), \quad \katou{ 0 \leq \theta < 2 \pi}, \label{eq:kj_mix_density} 
\end{align}
where $m \in \mathbb{N}$ is the number of the components of the mixture and $0 < \pi_1,\ldots,\pi_m < 1$ are the weights of the components satisfying $\sum_{k=1}^m \pi_k =1 $.
The parameters of each component, namely, $\mu_k$, $\gamma_k$, $\rho_k$ and $\lambda_k$, are defined as in (\ref{eq:kj_density}).
}

\subsection{Components of the mixture}
%Each component of the mixture (\ref{eq:kj_mix_density}) 
%The parameter $\mu$ and $\gamma$ control the location and concentration of the distribution, respectively. 
%The other two parameters, $\lambda$ and $\rho$, regulate the skewness and kurtosis.
%The distribution is unimodal if $\gamma>0$ and uniform if $\gamma=0$.
\katott{Here we introduce three important benefits of the Kato--Jones distribution (\ref{eq:kj_density}), which motivate us to use this model as the components of the proposed mixture.}

First, \chris{the} Kato--Jones distribution is a flexible unimodal distribution that can provide a wide range of skewness and peakedness.
As can be seen in Fig.\ 1(b) and (c) of \citet{kat15}, the degrees of skewness and peakedness of \chris{the} Kato--Jones distribution can be widely controlled by adjusting the parameters.
Unimodality is also important to \katot{interpret each component} of the mixture.

The second benefit is that each parameter of \chris{the} Kato--Jones distribution is easily interpretable. The parameter $\mu$ controls the location or mean direction, while $\gamma$ regulates the concentration or the mean resultant length.
The parameters $\lambda$ and $\rho$ influence the skewness and kurtosis of the model.
As shown in Lemma 1 of \citet{kat15}, the circular skewness and kurtosis of \citet{bat} \chris{for the} Kato--Jones distribution (\ref{eq:kj_density}) are given by $\bar{\beta}_2=\rho \gamma \sin \lambda$ and $\bar{\alpha}_2=\rho \gamma \cos \lambda$, respectively.
\katos{Note that kurtosis is related to the peakedness of the density \citep[see][Fig.\ 1(b)]{kat15}.
The} Kato--Jones distribution can be reparametrized \katott{to have the following density with the parameters $\mu, \gamma, \bar{\alpha}_2$ and $\bar{\beta}_2$:  %implying that all the parameters of each cluster of our mixture can be clearly interpreted.
\begin{align*}
\lefteqn{ g_{\rm KJ}^*(\theta; \mu,\gamma, \bar\alpha_2, \bar\beta_2) } \hspace{1cm} \\
& = \frac{1}{2\pi} \Biggl[ 1  + 2 \gamma^2  \frac{\gamma \cos 
(\theta - \mu) - \bar\alpha_2 }{\gamma^2 + \bar\alpha_2^2  + \bar\beta_2^2 
- 2 
\gamma 
\{ \bar\alpha_2 \cos (\theta-\mu) + \bar\beta_2 \sin (\theta -\mu) \} } 
\Biggr].
\end{align*}
Consequently, the density of the proposed mixture (\ref{eq:kj_mix_density}) can be written as
\begin{equation}
f(\theta) = \sum_{k=1}^m \pi_k \, g_{\rm KJ}^*(\theta; \mu_k,\gamma_k, \bar\alpha_{2k}, \bar\beta_{2k}), \label{eq:repa_density}
\end{equation}
where $\katou{0 \leq \mu_k < 2 \pi,} \ 0 \leq \gamma_k <1,$ and  
 $(\bar\alpha_{2k},\bar\beta_{2k}) \neq (\gamma_k,0)$ 
satisfy $(\bar\alpha_{2k} - \gamma_k^2)^2 + \bar\beta_{2k}^2 \leq \gamma_k^2  (1-\gamma_k)^2.$
%$\bar{\beta}_{2k}=\rho_k \gamma_k \sin \lambda_k$ and $\bar{\alpha}_{2k}=\rho_k \gamma_k \cos \lambda_k$ are the circular skewness and kurtosis of the $k$-th component, respectively.
With this representation of the density, all the parameters of each component of our mixture can be clearly interpreted;
the parameters $\mu_k$, $\gamma_k$, $\bar\alpha_{2k}$ and $\bar\beta_{2k}$ control the mean direction, mean resultant length, circular kurtosis and circular skewness of the $k$th component of the mixture, respectively.
}

Finally, parameter estimation for \chris{the} Kato--Jones distribution is straightforward by both method of moments and maximum likelihood.
This is mainly due to the fact that the trigonometric moments and probability density function of \chris{the} Kato--Jones distribution (\ref{eq:kj_density}) can be expressed in simple and closed form.
In general, parameter estimation for \katot{mixture models is involved.}
However the parameters of our mixture model (\ref{eq:kj_mix_density}) can be estimated with no great computational cost because of the simplicity of estimation for \chris{the} Kato--Jones distribution. 

The authors are not aware of other distributions which have all the three benefits above.
Note also that the use of \chris{the} Kato--Jones distribution is recommended in recent review papers of \citet{ley21} and \citet{pew} as a flexible distribution on the circle.

\katosss{
\subsection{Skewness of mixtures of symmetric distributions}
In general, mixtures of symmetric distributions have skewed \chris{density shapes}.
Then a natural question is whether it is necessary to use \katot{the mixtures of} Kato--Jones distributions rather than mixtures of symmetric distributions for the modeling of skewed data.
The following theorem provides an answer to this question.
\begin{theorem} \label{thm:skewness}
\katott{
Let $g_{\rm KJ}$ be the density of the Kato--Jones distribution (\ref{eq:kj_density}). Assume that $h_k (\theta)$ $(k=1,\ldots,m)$ is the density of a distribution on the circle which is symmetric about $\theta=\mu_k \,(\in \katou{[0,2 \pi)})$ and satisfies $\katou{\int_{0}^{2 \pi}} \cos \{ p (\theta-\mu_k)\} h_k(\theta) d\theta >0$ for any $p \in \mathbb{N}$.
Suppose $g_{\rm KJ}$ can be expressed as $g_{\rm KJ}(\theta) = \sum_{k=1}^m \pi_k h_k (\theta) $, where $\sum_{k=1}^m \pi_k =1$ and $\pi_k >0$.
Assume $\mu_k-\mu-\lambda$ is a rational number, i.e., $\mu_k-\mu-\lambda=a_k/b_k$ for some $a_k \in \mathbb{Z}$, $b_k=\mathbb{N}$. 
Then $g_{\rm KJ}$ is a symmetric density with $\lambda =0$.}
\end{theorem}
See Appendix D.1 of Supplementary Material for the proof.
%See Appendix \ref{sec:proof_skewness} for the proof.

This theorem suggests that \chris{an} asymmetric Kato--Jones distribution essentially has a different skewed shape \katott{from mixture of symmetric distributions under certain conditions.}
Our experiences suggest that it is difficult to model strong skewness using mixtures of symmetric distributions.
Indeed, as will be discussed later, mixtures of Kato--Jones distributions provide a much better fit to our data than those of the von Mises distributions because of the lack of fit in the skewed shape between 5 a.m. and 9 a.m.
Also it does not seem to be appropriate to use a mixture of symmetric distributions as a single component skewed distribution because a mixture of symmetric distributions is generally multimodal and therefore its interpretation is \katot{difficult}.

Note that many well-known distributions satisfy the assumption \chris{$\katou{\int_{0}^{2 \pi}} \cos \{ p (\theta-\mu_j)\} h_j(\theta) d\theta >0$ for any $p \in \mathbb{N}$} in Theorem \ref{thm:skewness}, including the von Mises, wrapped Cauchy and wrapped normal distributions.
\katott{
In addition a stronger statement can be made for mixtures of the wrapped Cauchy distributions as follows.
\begin{theorem} \label{thm:skewness_wc}
    Let $g_{\rm KJ}$ be the Kato--Jones density (\ref{eq:kj_density}) with $\gamma < \rho$.
    Suppose that $h_{{\rm WC},k} (\theta)$ $(k=1,\ldots,m)$ is the wrapped Cauchy distribution with density
    $$
    h_{{\rm WC},k} (\theta) = \frac{1}{2\pi} \frac{1-\rho_k^2}{1+\rho_k^2- 2 \rho_k \cos (\theta - \mu_k)}, \quad \katou{ 0 \leq \theta < 2 \pi},
    $$
    where $\katou{0 \leq \mu_k < 2 \pi }$, $ 0 \leq \rho_k <1 $, and $(\mu_k,\rho_k) \neq (\mu_{\ell},\rho_{\ell})$ for $k \neq \ell$.
    Assume that $g_{\rm KJ}(\theta) = \sum_{k=1}^m \pi_k h_{{\rm WC},k} (\theta) $, where $\sum_{k=1}^m \pi_k =1$ and $\pi_k >0$.
    Then $g_{\rm KJ}(\theta)$ is expressed as the following density of the symmetric two-component mixture
    \begin{equation}
    g_{\rm KJ}(\theta) = \frac{\gamma}{\rho} \frac{1}{2\pi} \frac{1-\rho}{1+\rho^2 - 2 \rho \cos (\theta - \mu)} + \left(1 - \frac{\gamma}{\rho} \right) \frac{1}{2\pi}, \quad \katou{0 \leq \theta < 2\pi}. \label{eq:two_wc}
    \end{equation}
\end{theorem}
See Appendix D.2 of Supplementary Material for the proof.
This theorem implies that any asymmetric case of the Kato--Jones distribution can not be expressed as mixtures of the wrapped Cauchy distributions.
}

}

\section{Parameter Estimation} \label{sec:estimation}
We consider parameter estimation for the proposed mixture (\ref{eq:kj_mix_density}) by maximum likelihood and a modified method of moments.
A key reparametrization is carried out to circumvent a problem in parameter estimation.
In our data analysis, the maximum likelihood estimator is mainly discussed because of its efficiency.
However the calculation of the modified method of moments estimate is faster and this estimate is used as an initial value of our algorithm for maximum likelihood estimation.

\subsection{Reparametrization}
Before we proceed to parameter estimation for the mixture (\ref{eq:kj_mix_density}), we \kator{reparametrize} \chris{it} to \katot{avoid potential problems} associated with parameter estimation.
\kator{First we provide alternative expressions for the densities of Kato--Jones distributions and their mixture.
See Appendix D.3 of Supplementary Material for the proof.
%See Appendix \ref{sec:proof_reparametrization} for the proof.
\begin{prop} \label{prop:reparametrization}
The following hold for the densities of Kato--Jones distributions (\ref{eq:kj_density}) and their mixture (\ref{eq:kj_mix_density}).
\begin{enumerate}[(i)]
    \item
The probability density function of Kato--Jones distribution (\ref{eq:kj_density}) can be expressed as
\begin{align}
\lefteqn{ g_{\rm KJ}(\theta;\mu,\gamma,\rho,\lambda) } \hspace{0.5cm} \nonumber \\
		& =  \pi' \cdot \frac{1}{2\pi} \left\{ 1 + 2 \bar{\gamma} \, \frac{\cos 
		(\theta-\mu)  - \rho \cos \lambda}{1+\rho^2-2 \rho \cos 
		(\theta-\mu-\lambda)} \right\} + \left( 1- \pi' \right) \frac{1}{2\pi}, \label{eq:kj_density2}
\end{align}
where $\pi'=\gamma / \bar{\gamma} \, (\in (0,1])$ and $\bar{\gamma} = (1-\rho^2)/\{2(1-\rho \cos \lambda )\} $ is the upper bound of the range of $\gamma$ for given $\rho$ and $\lambda$.
\item An alternative expression for the density of the mixture (\ref{eq:kj_mix_density}) is
\begin{equation}
    \katov{
	f(\theta) = \frac{1}{2\pi} \sum_{k=1}^m \pi_k' \cdot g_{\rm KJ}(\theta;\mu_k,\bar{\gamma}_k,\rho_k ,\lambda_k )  + \frac{\pi'_{m+1}}{2\pi}, }\label{eq:kj_mix_density2}
\end{equation}
\katov{where $g_{\rm KJ}$ is as in (\ref{eq:kj_density2}),} $\pi_k'= \pi_k \gamma_k / \bar{\gamma}_{k} $ $(k=1,\ldots,m)$, $\pi_{m+1}'=1-\sum_{k=1}^m \pi_k'$, and $\bar{\gamma}_{k} = (1-\rho_k^2)/\{2(1-\rho_k \cos \lambda_k )\} $.
\end{enumerate}
\end{prop}
}

\kator{Proposition \ref{prop:reparametrization} implies that} the proposed mixture (\ref{eq:kj_mix_density}) can be viewed as a mixture of Kato--Jones \katos{submodels} $g_{\rm KJ}(\mu_k,\bar{\gamma}_k,\lambda_k,\rho_k)$ and a uniform distribution.
The expression (\ref{eq:kj_mix_density2}) \katos{suggests} that $\pi_k$ and $\gamma_k$ cannot be uniquely determined in parameter estimation because the information \chris{on} both parameters is contained in the single parameter $\pi_k'$.
\katot{This implies that the parameters of the proposed mixtures reduce to $(\mu_1,\ldots,\mu_m,\rho_1,\ldots,\rho_m,\lambda_1,\ldots,\lambda_m,\pi'_1,\ldots,\pi'_{m+1})$.}
\katott{With this parametrization, it is possible to prove the identifiability of the proposed family of mixtures.
See Appendix D.4 for the proof.
}

\katott{
\begin{theorem} \label{thm:identifiability}
Consider a family of mixture models with the density (\ref{eq:kj_mix_density2}),
%\begin{equation}
%	f(\theta) = \frac{1}{2\pi} \sum_{k=1}^m \pi_k' \left\{ 1 + 2 \bar{\gamma}_k \, \frac{\cos (\theta-\mu_k)  - \rho_k \cos \lambda_k}{1+\rho_k^2-2 \rho_k \cos (\theta-\mu_k-\lambda_k )} \right\} + \frac{\pi'_{m+1}}{2\pi}, \label{eq:kj_mix_density3}
%\end{equation}
where $\mu_k,\rho_k,\lambda_k $ and $\pi_{\ell}'$ are the parameters of the family and $\bar{\gamma}_{k} = (1-\rho_k^2)/\{2(1-\rho_k \cos \lambda_k )\} $ $(k=1,\ldots,m, \ \ell=1,\ldots,m+1)$.
Define the parameter space of the family (\ref{eq:kj_mix_density2}) by
\begin{align}
\begin{split} \label{eq:omega}
\Omega = & \biggl\{ (\mu_1,\ldots,\mu_m,\rho_1,\ldots,\rho_m,\lambda_1,\ldots,\lambda_m,\pi'_1,\ldots,\pi'_{m+1}) \, \Bigl| \,   \\
& \quad \mu_k,\lambda_k \in \katou{[0,2 \pi)}, \ \rho_k \in (0,1),\ \rho_k > \rho_{\ell}, \ 1 \leq k < \ell \leq m, \\
& \quad \pi'_q> 0 , \ \sum_{q=1}^{m+1} \pi'_{q} =1 , \   q=1, \ldots, m+1 , \ m \in \mathbb{N} \biggr\}.
\end{split}
\end{align}
Then the family of the mixtures (\ref{eq:kj_mix_density2}) is identifiable under $\Omega$. 
\end{theorem}
}

\katos{Therefore, instead of estimating the parameters of (\ref{eq:kj_mix_density}) directly, we estimate the reparametrized mixture (\ref{eq:kj_mix_density2}) to ensure the identifiability of the model.

In order to interpret the estimated model, we consider \katot{the following two approaches:}
\katot{
\begin{enumerate}
\item[(a)] The first approach is to interpret the reparametrized mixture (\ref{eq:kj_mix_density2}) directly.
\item[(b)] Alternatively, we recover the original parameters, (\ref{eq:kj_mix_density}) or (\ref{eq:repa_density}), from the reparametrized ones (\ref{eq:kj_mix_density2}) and interpret the mixture with the recovered parameters.
\end{enumerate}
}

For achieving the recovery of the parameters in approach (b), it is necessary to add an additional assumption on the parameters to \katot{allocate} the mixing proportion, $\pi_{m+1}'$, of the uniform component to the other mixing proportions, \katos{$\{\pi_k'\}_{k=1}^m$.}
To this end, we propose the assumption $\pi_k= \pi_k' / ( 1-\pi_{m+1}')$ between the original and reparametrized parameters.
Although it is mathematically impossible to find which cluster observations from the uniform distribution in (\ref{eq:kj_mix_density2}) belong to, it seems natural to assume that the ratio of the uniform observations in the $k$-th cluster is proportional to that of the observations from Kato--Jones distribution $g_{\rm KJ}(\mu_k,\bar{\gamma}_k, \lambda_k, \rho_k)$ in the $k$-th cluster.

%To this end,} we need an additional condition on the parameters to ensure the identifiability of the mixture (\ref{eq:kj_mix_density}) (apart from swapping the parameters of each component).
%The mixture (\ref{eq:kj_mix_density}) is not identifiable for $p \geq 2$ in general because $\pi_h$ and $\gamma_h$ are not uniquely determined.

%In our paper, we propose the following approach by which} the flexibility of the mixture (\ref{eq:kj_mix_density}) is maintained.
% 
% However the expression (\ref{eq:kj_mix_density2}) also provides a solution to this problem since we can avoid
%First, we reparametrize $\{ \pi_k,\gamma_k \}_{k=1}^m$ to $\{\pi_k'\}_{k=1}^{m+1}$ via $\pi_k'= \pi_k \gamma_k / \bar{\gamma}_k $ and estimate the parameters of the reparametrized model (\ref{eq:kj_mix_density2}).
%Then, in order to recover the original parameters $\{ \pi_k, \gamma_k \}$ from the reprarametrized ones $\{ \pi_k'\}$, %we need an additional assumption on the parameters.

Summarizing the results above, we make the following steps to estimate the parameters of the mixture (\ref{eq:kj_density}):
\begin{enumerate}[i)]
	\item Estimate the parameters $\{\mu_k,\rho_k,\lambda_k,\pi_k'\}$ of the mixture (\ref{eq:kj_mix_density2}) to obtain their estimates $\{\hat{\mu}_k,\hat{\rho}_k,\hat{\lambda}_k,\katot{\hat{\pi}'_k}\}$. \label{item:step1}
	\item Recover the estimates of the original parameters $\{ \hat{\mu_k}, \hat{\gamma}_k, \hat{\rho}_k, \hat{\lambda}_k, \katot{\hat{\pi}_k} \}$ via $\hat{\pi}_k=\hat{\pi}'_k /(1-\hat{\pi}'_{m+1})$ and $\hat{\gamma}_k = \hat{\pi}_k' \hat{\bar{\gamma}}_{k}/ \hat{\pi}_k$. \label{item:step2}
\end{enumerate}
Then we interpret the reparametrized mixture (\ref{eq:kj_mix_density2}) estimated in Step \ref{item:step1}) as in approach (a) or the \katoss{original} mixture (\ref{eq:kj_mix_density}) estimated in Step \ref{item:step2}) as in approach (b).
Note that the mixture estimated via approach (a) or (b) has the same flexibility as the original mixture (\ref{eq:kj_mix_density}) although the former mixture has $m-1 (=(5m-1)-4m)$ \chris{fewer} parameters than the latter mixture.
In this paper we mainly discuss the interpretation of the mixture (\ref{eq:kj_mix_density}) estimated in Step \ref{item:step2}).

%Another possible solution to ensure the identifiability of the mixture (\ref{eq:kj_mix_density}) is to use the asymmetric three-parameter Kato--Jones distributions (2015, equation (7)) as the  components of the mixture (\ref{eq:kj_mix_density}).
%This can be easily done by imposing $\gamma_h= \rho_h \cos \lambda_h$.
 % is maintained in the reparametrized mixture (\ref{eq:kj_mix_density2}) with this condition on the parameters.

In order to achieve the identifiability of the Kato--Jones mixtures \katott{ with the original parametrization (\ref{eq:kj_mix_density})}, an alternative assumption on the parameters is to adopt the same assumption as the asymmetric three-parameter Kato--Jones distributions \citep[equation (7)]{kat15} for each component of the mixture (\ref{eq:kj_mix_density}).
This can be easily done by imposing $\gamma_k= \rho_k \cos \lambda_k$ in (\ref{eq:kj_mix_density}).
However the degree of \chris{kurtosis}, which influences the \chris{peakedness}, \katos{of each component} cannot be regulated for this mixture.
On the other hand, the reparametrized mixture (\ref{eq:kj_mix_density2}) achieves much greater flexibility than the mixture of three-parameter asymmetric Kato--Jones submodels although the reparametrized mixture (\ref{eq:kj_mix_density2}) has only one more parameter than \chris{latter mixture.}}
% We mainly discuss parameter estimation for the reprametrized model (\ref{eq:kj_mix_density2}) because the reprametrized mixture (\ref{eq:kj_mix_density2})

%Note that the problem associated with the identifiability of the uniform component is not unique to the mixture of Kato--Joned distributions (\ref{eq:kj_mix_density}).

%The expression (\ref{eq:kj_mix_density2})

%Then the fun

\subsection{Modified method of moments estimation}
\label{sec:ETM}
Throughout this \katot{section}, let \katos{random variables} $\Theta_1,\ldots,\Theta_n$ be independent and identically distributed from the reparametrized mixture (\ref{eq:kj_mix_density2}).
Denote the parameters of the mixture (\ref{eq:kj_mix_density2}) by
$$
\bm{\Psi}=(\mu_1,\ldots,\mu_m,\rho_1,\ldots,\rho_m,\lambda_1,\ldots,\lambda_m,\pi'_1,\ldots,\pi'_{m+1}).
$$

First we discuss the modified method of moments estimation based on trigonometric moments.
It is straightforward from \citet[equation (2)]{kat15} \chris{to see} that \katot{the $p$th trigonometric moment of $\Theta_j$ is} given by
\begin{equation}
	E (e^{i p \Theta_j}) = \sum_{k=1}^m \pi'_k \bar{\gamma}_{k} \,(\rho_k e^{i \lambda_k})^{-1}  \left\{ \rho_k e^{i (\mu_k + \lambda_k )} 
	\right\}^p, \quad p \in \mathbb{N}, \label{eq:moments}
\end{equation}
where $i$ is the imaginary unit.
%where $\alpha_h = -\lambda_h$,  $\beta_h = \gamma_h / \rho_h$ and $\eta_h = \mu_h + \lambda_h$ are reparametrized parameters for convenience in presentation.
Ordinary method of moments estimation based on \katot{the} trigonometric moments is obtained by equating the theoretical trigonometric moments (\ref{eq:moments}) to empirical ones, namely,
\begin{equation}
	\frac{1}{n} \sum_{j=1}^n e^{i p \Theta_j} =  \sum_{k=1}^m \pi'_k \bar{\gamma}_{k} \,(\rho_k e^{i \lambda_k})^{-1}  \left\{ \rho_k e^{i (\mu_k + \lambda_k )} 
	\right\}^p, \label{eq:method_moments}
\end{equation}
for some selected values of $p$.
However there is a problem associated with the ordinary method of moments estimation for the proposed mixture (\ref{eq:kj_mix_density}).
As is the case for $p=1$, the solutions to equation (\ref{eq:method_moments}) are not always within the range of $\lambda_k$ and $\rho_k$ \citep[see][Section 5.1]{kat15}.
%(ii) The second problem is that the parameters of the mixture (\ref{eq:kj_mix_density}) are not identifiable for $p \geq 2$.
In order to circumvent this problem, we consider the following function to evaluate the \kator{weighted} error between the empirical and theoretical trigonometric moments\chris{:}
\begin{align*}
{\rm ETM} (\bm{\Psi}) %} \hspace{1.5cm} \\
& \equiv \sum_{p=1}^q w(p) \left| \frac{1}{n} \sum_{j=1}^n e^{i p \Theta_j} -  E (e^{i p \Theta}) \right|^2 \nonumber \\
& =  \sum_{p=1}^q w(p) \left| \frac{1}{n} \sum_{j=1}^n e^{i p \Theta_j} - \sum_{k=1}^m \pi'_k \bar{\gamma}_{k} \,(\rho_k e^{i \lambda_k})^{-1}  \left\{ \rho_k e^{i (\mu_k + \lambda_k )} 
\right\}^p \right|^2,  %\label{eq:tme}
\end{align*}
where $q \in \mathbb{N}$ and $w(p)$ is a weight function.
%（加藤コメント：以前の原稿では、MEFを用いていましたが、こちらの方が意味がわかりやすいかと思い変更しました。）
%（加藤の備忘録：参考文献を引用する必要あり？）
This function can also be expressed as
\begin{align*}
 {\rm ETM} ( \bm{\Psi} )  %\hspace{0.5cm} \nonumber \\
	%\begin{split}
    = \ & \sum_{p=1}^q w(p) \Biggl[ \biggl\{ \frac1n \sum_{j=1}^n \cos p \Theta_j - \sum_{k=1}^m \pi'_k \bar{\gamma}_{k} \cdot \rho_k^{p-1} \cos \left( p \mu_k + (p-1) \lambda_k \right)  \biggr\}^2 \nonumber \\
		& +  \biggl\{ \frac1n \sum_{j=1}^n \sin p \Theta_j - \sum_{k=1}^m \pi'_k \bar{\gamma}_{k} \cdot \rho_k^{p-1} \sin \left( p \mu_k + (p-1) \lambda_k \right)  \biggr\}^2 \Biggr] . \label{eq:tme2}
	%\end{split} \label{eq:mef_r}
\end{align*}
Then we propose a modified method of moments estimator as the minimizer of ${\rm ETM}$, namely,
\begin{equation}
\hat{\bm{\Psi}} = \argmin_{\bm{\Psi} \in \Omega}  {\rm ETM} ( \bm{\Psi}),  \label{eq:mme}
\end{equation}
where $\Omega$ is the parameter space of $\bm{\Psi}$ defined in (\ref{eq:omega}).
If there exists a solution $\hat{\bm{\Psi}} \in \Omega$ to equation (\ref{eq:method_moments}) for $p=1,\ldots,q$, then ${\rm ETM} ( \hat{\bm{\Psi}})=0$ and $\hat{\bm{\Psi}}$ becomes the same as the ordinary method of moments estimate.
Therefore the proposed estimator (\ref{eq:mme}) is a generalization of the method of moments estimator.
An advantage of the proposed estimator is that the estimate always belongs to the parameter space.
For a single distribution $m=1$ with $w(1)=c$ and $w(p)=0 \ (p \geq 3)$, this estimator converges to the method of moments estimator of \citet{kat15} as $w(2) \rightarrow 0$.

%$\hat{\bm{\Psi}}$ becomes the same as the ordinary method of moments estimate as $n \rightarrow \infty$.
%Since we have if there exists a solution $\hat{\bm{\Psi}} \in \Omega$ to equation (\ref{eq:method_moments}) for $p=1,\ldots,q$,.

%In our case, this problem boils down to investigating if the minimization in the equation (\ref{eq:mme}) has the unique solution, and it appears difficult to prove this in general.
%However, as seen in Theorem \ref{thm:identifiability}, the reparametrized mixture (\ref{eq:kj_mix_density2}) is identifiable and, unlike general mixture models, we will not have a problem of estimating different parameters for the same distribution. 
%Also, our experiments suggest that the modified method of moments estimation provides reasonable estimates in most cases.

\katorrr{The proposed estimator (\ref{eq:mme}) is somewhat related to the estimator for the stable distribution on $\mathbb{R}$ discussed in \citet{pre72}.
Both estimators are $M$-estimators based on the \katos{square of \chris{the} absolute} difference between theoretical and empirical functions related to the characteristic functions.
However our estimator is essentially different from \katoss{the estimator of \citet{pre72}} in the sense that our estimator is for a different distribution on a different manifold and avoids \chris{an} integral over the infinite interval of the argument of the function.}

%Since the proposed estimator (\ref{eq:mme}) is an $M$-estimator, the consistency of the estimator immediately follows from general theory \citep[see][Section 4.2c]{ham}.
%In addition, since ${\rm ETM}$ is differentiable with respect to its parameters, the asymptotic variance of the estimator can also be evaluated.

The result of the estimation (\ref{eq:mme})  depends on the choice of $q$ and $w(p)$. 
Since the mixture (\ref{eq:kj_mix_density2}) has essentially $4m$ free parameters, there are potentially multiple solutions to the equation ${\rm ETM} ( \hat{\bm{\Psi}})=0$ for $q < 2m$.
To avoid this potential problem, it is recommended to assume $q \geq 2m$ and, in our analysis, we assume that $q=2m$.
%（加藤コメント：$q$はいくつに設定しましたか？）
As for the choice of the weight function $w(p)$, we adopt $w(p)=c^p (0<c<1)$ in our data analysis to place much more importance on the low-order \katos{trigonometric} moments than on the high-order ones.
This choice is made in a somewhat similar spirit to the method of moments estimation \katos{of \citet{kat15} in which a couple of estimates are determined only by the first trigonometric moment.}

\katott{
It is known that, under certain regularity conditions, consistency holds for $M$-estimators.
The modified method of moments estimator (\ref{eq:mme}) defined for the identifiable model given in Theorem \ref{thm:identifiability} satisfies all the regularity conditions for consistency given in Case A of Section 6.2 of \cite{hub}.
%These regularity conditions of the general $M$-estimators can be found, for example, in Sections 6.2 and 6.3 of \cite{hub}.
One of the regularity conditions for consistency is the following convergence:
$$
    {\rm ETM}(\hat{\bm{\Psi}}) - \inf_{\bm{\Psi} \in \Omega } {\rm ETM}(\bm{\Psi}) \longrightarrow 0 \quad \mbox{almost surely \ as } n \rightarrow \infty. \label{eq:convergence}
$$
This convergence does not hold if mixture models are not identifiable or objective functions for minimization have multiple solutions.
However, as seen in Theorem \ref{thm:identifiability}, our reparametrized mixture (\ref{eq:kj_mix_density2}) is identifiable.
In addition, if $q$ is appropriately selected, the equation ${\rm ETM}(\hat{\bm{\Psi}})=0$ has a unique solution (which is not necessarily within the parameter space), and the law of large numbers guarantees that the solution almost surely converges to the true value in the parameter space.
The other regularity conditions for consistency given in \citet[Section 6.2, (A.1)--(A.5)]{hub} are satisfied for our simple objective function and parameter space.

As for the asymptotic normality of the estimator (\ref{eq:mme}), it does not seem straightforward to prove the regularity conditions of \citet[Section 6.3]{hub}.
Specifically, we need to prove that $\partial 
 E ({\rm ETM} (\bm{\Psi}) ) / \partial \bm{\Psi}$ has a nonsingular derivative matrix as assumed in Corollary 3.2 of \citet{hub}, and this would be future work. 
}

\subsection{Maximum likelihood estimation}
Next we consider the maximum likelihood estimation.
The log-likelihood function for the sample $(\theta_1,\ldots,\theta_n)$ is given by
\begin{equation}
\ell ( \bm{\Psi})  %\hspace{0.5cm} %\nonumber \\
 = C + \sum_{j=1}^n \log \left[ \sum_{k=1}^m \pi'_k \left\{ 1 + 2 \bar{\gamma}_{k} \, \frac{\cos (\theta_j-\mu_k)  - \rho_k \cos \lambda_k}{1+\rho_k^2-2 \rho_k \cos (\theta-\mu_k-\lambda_k )} \right\} + \pi'_{m+1} \right]\kator{,} \label{eq:likelihood}
\end{equation}
\kator{where $C = -n \log (2 \pi)$.}
As is the case for $p=1$, there \chris{is no} closed-form expression for the maximum likelihood estimator \katot{for general $p$.}
Therefore we consider a numerical algorithm to estimate the maximum likelihood estimate of the mixture (\ref{eq:kj_mix_density2}).
We apply the EM algorithm to estimate the mixing proportions of the mixture (e.g., McLachlan and Krishnan, 2008, Section 1.4.3).
%Let $Z_1,\ldots,Z_n$ are independent and identically distributed 
\chris{The} following algorithm is established:
\begin{algo} \quad \label{alg:em}
\begin{enumerate}[Step 1:]
	\item Take an initial value $\bm{\Psi}^{(0)}$, where
	$$
	\bm{\Psi}^{(0)}=(\mu_1^{(0)},\ldots,\mu_m^{(0)},\rho_1^{(0)},\ldots,\rho_m^{(0)},\lambda_1^{(0)},\ldots,\lambda_m^{(0)},{\pi'}_1^{(0)},\ldots,{\pi'}^{(0)}_{m+1}).
	$$
	\item For $r=1,\ldots,N$, compute the following until the value of $\bm{\Psi}^{(N)}$ is virtually unchanged from $\bm{\Psi}^{(N-1)}$:
	$$
	{\pi'}_k^{(r)} = \frac{1}{n} \sum_{j=1}^n w_{k j}^{(r-1)}, \quad {\pi'}_{m+1}^{(r)} = 1 - \sum_{k=1}^m {\pi'}_k^{(r)},
	$$
	%Then  the parameters of each component of the mixture by
	\begin{equation}
		\left( \mu_k^{(r)}, \lambda_k^{(r)} , \rho_k^{(r)} \right) = \underset{(\mu_k,\lambda_k,\rho_k) }{\mathrm{argmax}} \left[ \sum_{j=1}^n  w_{k j}^{(r-1)} \log \left\{ g_{\rm KJ} (\theta_j ; \mu_k , \bar{\gamma}_{k}, \lambda_k ,  \rho_k ) \right\} \right], \label{eq:em_max}
	\end{equation}
where $k=1, \ldots, m$ and
$$
w_{kj}^{(r-1)} = \frac{{\pi'}_k^{(r-1)} g_{\rm KJ}(\theta_j; \mu_{k}^{(r-1)},\bar{\gamma}_{k}^{(r-1)}, \lambda_{k}^{(r-1)}, \rho_k^{(r-1)})}{ \sum_{h=1}^m {\pi'}_h^{(r-1)} g_{\rm KJ}(\theta_j; \mu_{h}^{(r-1)},\bar{\gamma}_{h}^{(r-1)}, \lambda_h^{(r-1)}, \rho_h^{(r-1)}) + {\pi'}_{m+1}^{(r-1)}/(2\pi) }.
$$

\item Record $\hat{\bm{\Psi}}^{(N)}$ as the maximum likelihood estimate of $\bm{\Psi}$.
\end{enumerate}
\end{algo}

An advantage of this algorithm is that the values of $\{ {\pi'}_k^{(r)} \}$ can be expressed in closed form.
In addition, although it \katos{is} necessary to calculate $( \mu_k^{(r)}, \lambda_k^{(r)} , \rho_k^{(r)} )$ numerically in (\ref{eq:em_max}), this maximization is essentially weighted maximum likelihood estimation for a single Kato--Jones distribution and can be done in a similar manner as in \citet[Section 5.2]{kat15}.

\katot{Our experiments suggest that the modified method of moments estimates (\ref{eq:mme}) provide useful initial values} of the algorithm.
However it is advisable to try multiple starting values to ensure the global maximum of the likelihood function.
\katot{The modified method of moments estimation} is faster than the maximum likelihood estimation.
There is\chris{, however,} no great difficulty in implementing maximum likelihood estimation using the algorithm above.

\katott{
Consistency and asymptotic normality hold for the maximum likelihood estimator under certain regularity conditions.
One of the regularity conditions for consistency given in \citet[(A1)]{miy20} is the boundedness of the likelihood function.
As the likelihood function (\ref{eq:likelihood}) with certain conditions on the parameters tends to $-\infty$ as $\rho_k \rightarrow 1$, we need to replace the range of $\rho_k$ of $\Omega$ in (\ref{eq:omega}) by $0 < \rho \leq 1-\varepsilon$ for a given small value of $\varepsilon >0$.
With this modification, the maximum likelihood estimator is consistent in the sense of Theorem 1 of \citet{miy20}.
In order to prove the asymptotic normality for the maximum likelihood estimator of our model, we need to show that the Fisher information matrix of $\bm{\Psi}$ is positive definite \citep[Condition 2]{red84} and that is our future work.
}

In our data analysis, we mainly discuss the maximum likelihood estimate because the maximum likelihood \katott{estimator has lower variance than the method of moments estimator (\ref{eq:mme}) for a large sample size as demonstrated in a simulation study in Section \ref{sec:simulation}.} %such as model selection via Akaike information criterion and log-likelihood ratio tests.

\katot{
\section{\kotan{Simulation study}} \label{sec:simulation}
}

We compare the performance of the modified method of moments estimation and the maximum likelihood estimation by EM algorithm via a Monte Carlo simulation study.
Random samples of sizes $n=50, 100, 500, 1000, $ and $5000$ were generated from the estimated distribution whose parameters are shown in Table \ref{tab:em} (b).
For each sample size, $r=2000$ samples were generated.

For comparison of the two estimators, we employ the generalized mean squared error which is defined by $\det{(\Sigma)}$, where $\Sigma=E\{(\hat{\xi}-\xi)(\hat{\xi}-\xi)^T\}$ and $\hat{\xi}$ is an estimator of \katot{$\xi=(\mu_1,\mu_2,\rho_1,\rho_2,\lambda_1,\lambda_2,\pi'_1,\pi'_2)^T$.}
An estimate of the generalized mean squared error is given by replacing $\Sigma$ by its sample analogue $\hat{\Sigma}=r^{-1}\sum^{r}_{j=1}{(\hat{\xi}_j-\xi)(\hat{\xi}_j-\xi)^T}$, where the $\hat{\xi}_j$'s\chris{,} $j=1,...,r$\chris{,} are the estimates from the $r$ simulation samples.
We \chris{also} consider the estimated relative generalized mean squared error of the estimate via the modified method of moments with respect to the maximum likelihood estimate, which is defined as
\begin{equation*}
	\widehat{\rm{RGMSE}}=\frac{\det{(\hat{\Sigma}_{MM}})}{\det{(\hat{\Sigma}_{ML}})}\chris{.}
	\label{eq:rgmse}
\end{equation*}
\chris{Here,} $\hat{\Sigma}_{MM}$ and $\hat{\Sigma}_{ML}$ are sample estimates of the mean squared error matrices of modified method of moments estimation and maximum likelihood estimation, respectively.

For numerical optimization of ETM, \chris{the} scipy.optimize.minimize package in Python with SLSQP method is employed \citep{jon01,kra88}.
This method is a kind of quasi-Newton method and enables us to solve the constrained optimization problem.
In this analysis, the optimization of ETM is terminated \chris{when} the difference \chris{in} the value of ETM between two successive steps is lower than $1\times10^{-10}$.
The initial values of $\mu_k$, $\rho_k$ and $\lambda_k$ are random samples from the uniform distribution on the intervals $[0,2\pi)$, $[0,1)$ and $[0,2\pi)$, respectively.
Also, those of $\pi_1'$, $\pi_2'$ and $\pi_3'$ are in $[0,1)$ and $\pi_1'+\pi_2'+\pi_3'=1$.
\kota{
For this sampling, two random values are drawn from the uniform distribution on $[0,1)$ and the smaller one is regarded as $r_1$ and the other as $r_2$.
Then, $\pi_1'=r_1$, $\pi_2'=r_2-r_1$, and $\pi_3'=1-r_2$ are calculated.}
Furthermore, the moment weight parameter $c$ in $w(p)$ is set to 0.9 because our experiments suggest that estimates with small $c$ are not stable with regard to fitting the data.

\kotar{100 estimates are obtained via this process, and the estimate whose value of ETM is smallest is set as the initial value for the maximum likelihood estimation.}
For numerical optimization \chris{in the} \katot{M-step, namely, Step 2 in Algorithm \ref{alg:em},} the same Python package as in the optimization of ETM is employed.
In this analysis, each M-step is terminated when the difference of the value of the objective function between two successive iterations is lower than $1\times10^{-6}$.
Also, \chris{the} whole EM algorithm is terminated when the difference of the log-likelihood function between two successive M-steps is lower than $n\times10^{-6}$.
In both estimates, it is assumed \chris{that} $\hat{\mu}_1<\hat{\mu}_2$.
%The conditions of the modified method of moments estimation \chris{are} the same \chriss{as} \chris{in} the previous section except for the number of initial values, which is set to 100 in this simulation.
%\chris{Maximum} likelihood estimation is achieved by the EM algorithm, whose initial value is the modified method of moments estimate and the threshold of termination for the EM algorithm is set to $n\times10^{-6}$.

\begin{table}
\caption{\label{tab:com} The values of $\det{(\hat{\Sigma}_{MM})}$, $\det{(\hat{\Sigma}_{ML})}$ and $\widehat{\rm{RGMSE}}$}
\centering
\fbox{
\begin{tabular}{c|ccccc}
$n$ & $n=50$ & $n=100$ & $n=500$ & $n=1000$ & $n=5000$ \\
\hline
$\det{(\hat{\Sigma}_{MM})}$ & $5.35\times10^{-9}$ & $7.00\times10^{-10}$ & $2.70\times10^{-13}$ & $4.25\times10^{-15}$ & $4.01\times10^{-20}$\\
$\det{(\hat{\Sigma}_{ML})}$ & $5.05\times10^{-8}$ & $6.41\times10^{-9}$ & $3.56\times10^{-12}$ & $1.33\times10^{-14}$ & $1.23\times10^{-22}$\\
$\widehat{\rm{RGMSE}}$ & 0.106 & 0.109 & 0.076 & 0.322 & 325.3\\
\end{tabular}}
\end{table}

\begin{figure}[t]
\centering
\includegraphics[clip,keepaspectratio,width=0.75\textwidth]{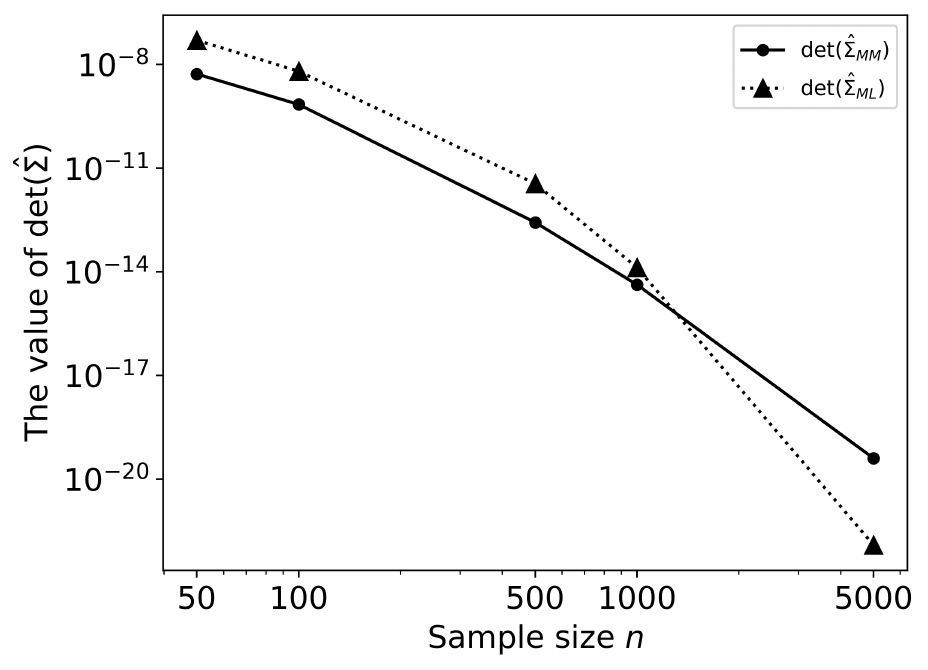}
\caption{\label{fig:simu} \katoss{Plot of the value of $\det{(\hat{\Sigma}_{MM})}$ (bold) and $\det{(\hat{\Sigma}_{ML})}$ (dotted) in simulation study.}}
\end{figure}

Table \ref{tab:com} shows the value of $\det{(\hat{\Sigma}_{MM})}$, $\det{(\hat{\Sigma}_{ML})}$ and $\widehat{\rm{RGMSE}}$ for each sample size.
\kotar{In addition, Fig.\ \ref{fig:simu} shows the trend of the change of the value of $\det{(\hat{\Sigma}_{MM})}$ and $\det{(\hat{\Sigma}_{ML})}$ in different sample sizes.}
As the sample size $n$ increases, the value of $\det{(\hat{\Sigma}_{MM})}$ and $\det{(\hat{\Sigma}_{ML})}$ \chris{decreases}.
\kotar{The value of $\det{(\hat{\Sigma}_{MM})}$ at $n=50$ is smaller than $\det{(\hat{\Sigma}_{ML})}$, however, the value of $\det{(\hat{\Sigma}_{MM})}$ at $n=5000$ is larger than $\det{(\hat{\Sigma}_{ML})}$ because the degree of decline in value with respect to sample size is greater for $\det{(\hat{\Sigma}_{ML})}$ than for $\det{(\hat{\Sigma}_{MM})}$.
Therefore, the values of $\widehat{\rm{RGMSE}}$ are smaller than 1 for $n\leq1000$ and larger for $n=5000$.}
\kotar{These results imply} modified moments estimation is preferable for estimation from a small number of samples \kotar{in terms of $\widehat{\rm{RGMSE}}$}.
On the other hand, maximum likelihood estimation is preferable for estimation from a large number of samples such as the data in this paper.
Note that modified moments estimation \chris{yields} not only an estimate but also a good initial value \chris{for} maximum likelihood estimation.
Although the value of $\widehat{\rm RGMSE}$ is relatively large for $n=5000$, the very small value of $\det{(\hat{\Sigma}_{MM})}$ for $n=5000$ indicates that the estimate via the modified moments estimation \chris{still has} reasonable performance as the initial value \chris{for} maximum likelihood estimation.
Therefore, modified moments estimation is useful to obtain the maximum likelihood estimates efficiently for a large sample size as well as small and medium sample sizes.
%Therefore, the the goodness of the modified moments estimation as a estimator cannot discuss by only the value of $\widehat{\rm{RGMSE}}$.
% n=5000とかでMMが意味ないわけではなく，初期値に使えるよねって話をする
%一方，MMはそれ単体で最終的な推定量とするのではなく，MLの初期値としていい値を提供するものとしての側面が強いと考えております．その際，両者ともに一致推定量であるためMLの初期値としての機能はある程度担保されるため，RGMSEの値でMMの推定量としての良し悪しは直接語れないのではないかなと思っております．

\kotar{It is noted that the degree of decline of $\det{(\hat{\Sigma})}$ with respect to sample size depends on the estimation method.
Therefore, for $n=500$, the value of $\widehat{\rm{RGMSE}}$ is slightly smaller than for $n=100$.
The tendency for $\widehat{\rm{RGMSE}}$ to not increase monotonically with respect to $n$ is also observed in the unimodal Kato--Jones distribution \citep[Supplementary Material]{kat15}.}

\quad

\section{Application} \label{sec:application}

\subsection{Application of modified method of moments estimation}
The proposed \katot{inferential methods are} applied to \katot{the traffic count data of interest}.
The number of samples is 1,121,262, as \chris{mentioned} in Section \ref{data}.
\kota{In this section,} the number of components is fixed to two because two \chris{main} peaks are recognized in the histogram (see Section \katot{6.2 for results for} other numbers of components).
The conditions of the modified method of moments estimation \chris{are} the same \chriss{as} \chris{in} the previous section.
\kotar{Note that \cite{nag23} in Japanese traffic engineering journal also analyzed traffic count data by applying the method proposed in the preprint version of this paper \citep{nag22}.
However, \cite{nag23} only discussed the ultimate results from the viewpoint of traffic engineering and did not discuss theoretical or statistical aspects of the method.
Moreover, they just focused on comparing the results of data from many counters in different locations, whereas this paper interprets the results from a single location in more detail.}

%the comparison between several  location of detction

\begin{figure}[t]
\centering
\includegraphics[clip,keepaspectratio,width=0.75\textwidth]{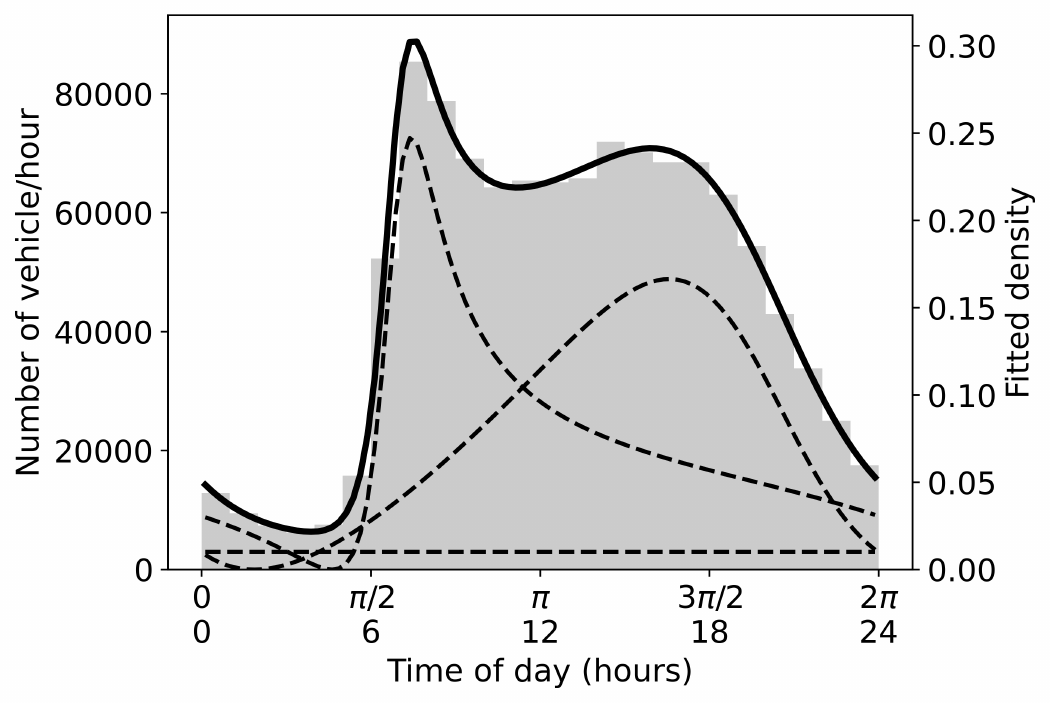}
\caption{\label{fig:best} \katoss{Plot of the densities of the reparametrized mixture (\ref{eq:kj_mix_density2}) (bold) and \chris{its} components (dashed) estimated via \kota{the modified method of moments estimation.}}}
\end{figure}

\begin{table}
\caption{\label{tab:best} \katoss{\chris{Parameter estimates} associated with the \kota{modified method of moments \katot{estimation}}.}}
\centering
\fbox{
\subtable[Initial value]{
\begin{minipage}{0.45\textwidth}
\begin{center}
\begin{tabular}{*{5}{c}}
$k$ & $\mu_k$ & $\rho_k$ & $\lambda_k$ & $\pi_k'$ \\
\hline
1 & 2.2755 & 0.5322 & 4.9979 & 0.0316 \\
2 & 3.9458 & 0.8131 & 2.0455 & 0.2331 \\
3 & & & & 0.7353 \\
\end{tabular}
\end{center}
\end{minipage}}
\subtable[Estimated parameter]{
\begin{minipage}{0.45\textwidth}
\begin{center}
\begin{tabular}{*{5}{c}}
$k$ & $\katot{\hat{\mu}}_k$ & \katot{$\hat{\rho}_k$} & \katot{$\hat{\lambda}_k$} & \katot{$\hat{\pi}_k'$} \\
\hline
1 & 2.7514 & 0.7322 & 5.3162 & 0.4543 \\
2 & 4.0106 & 0.1947 & 1.1589 & 0.4820 \\
3 & & & & 0.0637 \\
\end{tabular}
\end{center}
\end{minipage}}}
\end{table}

\begin{table}
\caption{\label{tab:bestm} \katot{Values} of the empirical and theoretical trigonometric \katoss{moments} \katot{(t.m.'s)} and \katot{squares of their differences in the best result.}}
\centering
\fbox{
\begin{tabular}{c|cccc}
& & empirical \katot{t.m.} & theoretical \katot{t.m.} & square of \katot{difference} \\
\hline
cosine & $p=1$ & $-3.29\times10^{-1}$ & $-3.29\times10^{-1}$ & $2.95\times10^{-14}$ \\
& $p=2$ & $-7.07\times10^{-2}$ & $-7.07\times10^{-2}$ & $5.64\times10^{-15}$ \\
& $p=3$ & $9.46\times10^{-2}$ & $9.46\times10^{-2}$ & $3.07\times10^{-15}$ \\
& $p=4$ & $-1.61\times10^{-2}$ & $-1.61\times10^{-2}$ & $8.35\times10^{-15}$ \\
\hline
sine &$p=1$ & $-1.23\times10^{-1}$ & $-1.23\times10^{-1}$ & $2.81\times10^{-14}$ \\
& $p=2$ & $-1.18\times10^{-1}$ & $-1.18\times10^{-1}$ & $1.21\times10^{-13}$ \\
& $p=3$ & $1.29\times10^{-2}$ & $1.29\times10^{-2}$ & $4.21\times10^{-16}$ \\
& $p=4$ & $6.97\times10^{-2}$ & $6.97\times10^{-2}$ & $7.43\times10^{-16}$ \\
\end{tabular}}
\end{table}

The optimization is carried out \katot{for} 10,000 initial values.
Fig.\ \ref{fig:best} and Table \ref{tab:best} show the best result \kotar{in the sense that the value of ETM is the smallest} among 10,000 estimations.
\katot{Table \ref{tab:bestm} provides the values of the empirical and theoretical trigonometric moments.
The solid line in Fig.\ \ref{fig:best} represents the estimated density of the mixture and the dashed lines refer to the estimated densities of the components.
\chris{The same convention will be used in} the following figures of fitted densities.}
The value of ETM is $7.06\times10^{-13}$ and that of \chris{the} log-likelihood function is \katos{$-$}1,840,939.
As shown in the right column of Table \ref{tab:bestm}, the difference between the values of empirical and theoretical trigonometric moments \katot{of all the degrees} are approximately 0, which means that \katot{this result} is very close to \katot{the result of ordinary method of moments estimation.}
\cyan{
Nonetheless, we generally need the proposed modified method of moments to ensure that the parameters are within their ranges, as explained in Section 4.2.
Actually, \katot{when the number of components $m$ is set to be greater than two, much larger values of ETM are obtained for some cases and the \katoss{ordinary method of moments} estimates are out of range for some parameters} (see Section \katos{6.2}).}
Also, the \katot{estimated densities} in Fig.\ \ref{fig:best} \chris{fit fairly} \katot{well} to the histogram.
The sharp and negatively skewed peak around $\pi/2$ is represented by the component \katot{with} $k=1$, \katot{while} the gentle peak around $3\pi/2$ is \katot{modelled} by the component \katot{with} $k=2$.

The computation time for 10,000 trials is about \kotar{14} sec.
\kota{The calculation was run on a \katot{Windows 10} computer with \kotar{an Intel Core i9-12900 processor running at 3.20 GHz using 64.0 GB of RAM.}}
Furthermore, about 25\% of 10,000 trials converge to \kotar{almost the same} value \kota{as} the best result.
\kotar{
In this case, 10,000 initial values were used to verify the performance of the modified method of moments estimation.
% 今回は性能を確かめるために10000個もの初期値を用いた
\kotar{The estimates that provide the smallest value of ETM} can be obtained from 25$\%$ of the initial values, which implies that estimation from a much smaller number of initial values (e.g., 100) would be sufficient.
}
% しかし，25%もの初期値から最適な推定量が得られるということは，もっと少ない(e.g. 100)初期値の個数から推定でも最適な推定量が得られることを意味する．
%We confirmed that various initial values reached the estimated parameters close to those in Table \ref{tab:best} \kota{and the initial value in Table \ref{tab:best} is just an example of such initial values.}

\kotar{Note that the modified method of moments estimation is carried out to obtain a good initial value for maximum likelihood estimation, which is more appropriate for such a large sample size, as discussed in Section 5.}

\subsection{Maximum likelihood estimation} \label{sec:mle}
\chris{The} EM algorithm is carried out by setting the initial values as the best estimates from \chris{the} modified method of moments to achieve the maximum likelihood estimation \kotar{efficiently}.
This choice of the initial values seems reasonable because of the large sample size of the data and the consistency of the modified method of moments estimator and maximum likelihood estimator.
\kotar{The conditions of the maximum likelihood estimation \chris{are} the same \chriss{as} \chris{in} the previous section except for the threshold for the termination of the EM algorithm, which is set to 1.}
This threshold is set by the following \chris{reasoning}.
Since the number of samples in this study is more than 1,000,000, the absolute value of \chris{the} log-likelihood function is expected to be \chris{of} that order of magnitude.
Thus, we regard the threshold 1 \chris{as} small enough to \chris{claim} convergence.

\begin{figure}[t]
\centering
\includegraphics[clip,keepaspectratio,width=0.75\textwidth]{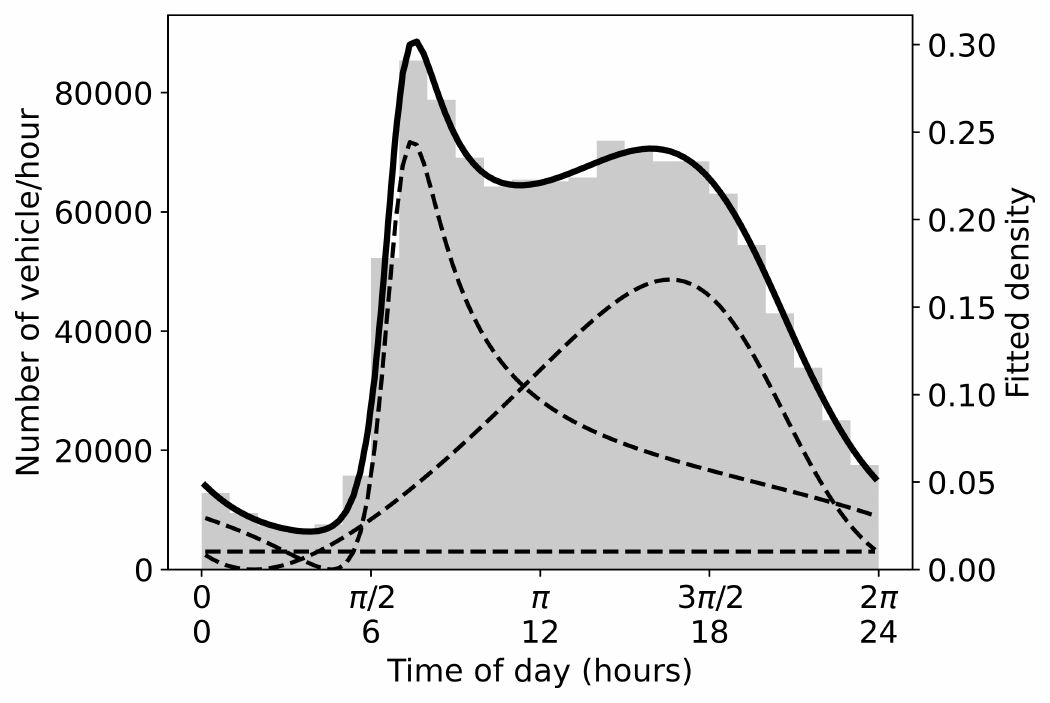}
\caption{\label{fig:em} \katoss{Plot of the densities of the reparametrized mixture (\ref{eq:kj_mix_density2}) (bold) and \chriss{its} components (dashed) estimated via the EM algorithm.}}
\end{figure}

\begin{table}
\caption{\label{tab:em} \chris{Parameter estimates} associated with the maximum likelihood estimation for the reparametrized mixture (\ref{eq:kj_mix_density2}) via \chris{the} EM algorithm.}
\centering
\fbox{
\begin{tabular}{*{5}{c}}
$k$ & $\hat{\mu}_k$ & $\hat{\rho}_k$ & $\hat{\lambda}_k$ & $\hat{\pi}_k'$ \\
\hline
1 & 2.7572 & 0.7266 & 5.3136 & 0.4536 \\
2 & 4.0107 & 0.1970 & 1.1895 & 0.4825 \\
3 & & & & 0.0639 \\
\end{tabular}}
\end{table}

The result of \chris{the} EM algorithm is shown in Fig.\ \ref{fig:em} and Table \ref{tab:em}.
It takes \chriss{just} two steps to terminate the EM algorithm.
\chris{The} value of ETM \chris{increases along with that of the log-likelihood} during the EM algorithm.
\chris{The} difference \katot{between the parameter estimates} by the modified method of moments and those by the maximum likelihood estimation \katot{is} not \chris{much}.
This implies \chris{that} \kota{the modified method of moments estimate is close to the \kotar{maximum likelihood estimate}, the estimate \chriss{produced} by another consistent estimator.}

\kotar{
In addition, the standard errors for the parameter estimates by the modified method of moments and those by the maximum likelihood estimation are calculated.
The standard errors for the parameter estimates by the modified method of moments are calculated by the nonparametric bootstrap method \citep{dav97}, and those by the maximum likelihood estimation are calculated by the Fisher information matrix for the log-likelihood function due to computation time.}

\begin{table}
\caption{\label{tab:stderr} \kotar{The value of standard errors for the parameter estimates by the modified method of moments and those by the maximum likelihood estimation.}}
\centering
\fbox{
\begin{tabular}{c|ccccc}
 & $k$ & $\hat{\mu}_k$ & $\hat{\rho}_k$ & $\hat{\lambda}_k$ & $\hat{\pi}_k'$ \\
\hline
the modified method of moments & 1 & $5.77\times10^{-2}$ & $2.46\times10^{-2}$ & $1.87\times10^{-1}$ & $8.91\times10^{-3}$\\
& 2 & $5.74\times10^{-2}$ & $2.47\times10^{-2}$ & $1.90\times10^{-1}$ & $5.92\times10^{-3}$\\
& 3 & & & & $6.59\times10^{-2}$\\
\hline
the maximum likelihood estimation & 1 & $6.36\times10^{-3}$ & $3.50\times10^{-3}$ & $2.88\times10^{-2}$ & $3.27\times10^{-3}$\\
& 2 & $6.82\times10^{-3}$ & $1.98\times10^{-3}$ & $9.03\times10^{-3}$ & $3.76\times10^{-3}$\\
& 3 & & & & $3.30\times10^{-3}$\\
\end{tabular}}
\end{table}

\kotar{The results are shown in Table \ref{tab:stderr}.
\katou{These results indicate that,} for all parameters, the values of standard error \katou{of the maximum likelihood estimates} are smaller than those of \katou{the modified method of moments estimates.
However the modified method of moments estimates generally exhibit satisfactory performance, supporting our idea of using these estimates as the initial values of the EM algorithm.
For both the modified method of moments and maximum likelihood estimates, the values of the standard errors of $\hat{\lambda}_k$ are greater than those of the other parameters.}
The values of standard error for the parameter estimates by the maximum likelihood estimation are in between the order of the second and third power of 10, and those by the modified method of moments are in between the order of the one and third power of 10.
%(事実を書いたのですが，ここからどういうことが言えるかがわからなくて考察に困っております，「ETMによる推定値はMLEよりもreliabilityに欠けるが，絶対的な評価では大きな問題はない」といったことを書けばいいのでしょうか？あと，そもそもこの話をどこに挿入するかも迷っております．)\katou{【加藤：「ETMによる推定値はMLEよりもreliabilityに欠けるが，絶対的な評価では大きな問題はない」と書くので良いと思います．そのような旨の考察の例を付け足してみました．場所はここで良いと思います．】}
}

As for the computation time, it takes \kotar{93} sec. for \chris{the} EM algorithm although only two M-steps are conducted in the calculation.
\kota{The calculation environment is the same as that of the previous \katoss{subsection}.}
Compared to the modified method of moments that takes only about \kotar{14} sec. for 10,000 trials, \chris{the} EM algorithm requires \chris{a} longer time.
Therefore, setting the initial value close to the maximum likelihood estimator is necessary to achieve the maximum likelihood estimation by \chris{the} EM algorithm in a feasible computation time.
The proposed modified method of moments is one possible solution for this purpose.

\kotar{Although two parameter estimation methods are introduced in this paper, the maximum likelihood estimation is preferred in terms of the mean squared error of the estimators and usage for model comparison.
The modified method of moments is useful to obtain a good initial value for the maximum likelihood estimation with a short computation time for large sample sizes.}

\begin{figure}[t]
    \begin{center}
			\includegraphics[clip, keepaspectratio, width=0.75\hsize]{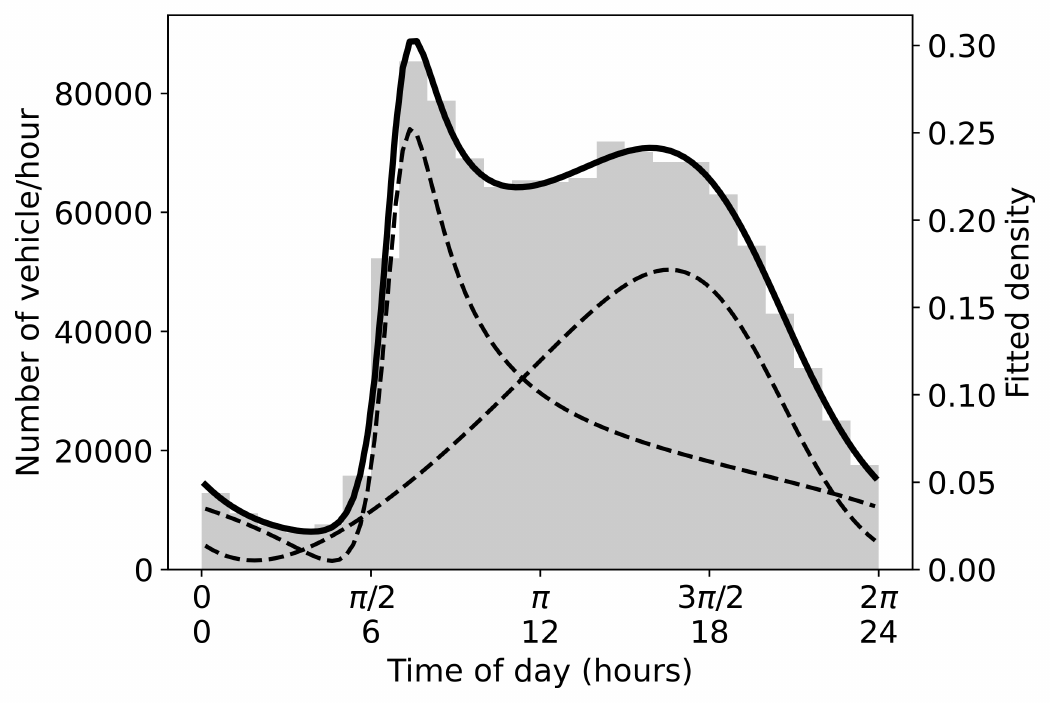}
    \end{center}
\caption{\label{fig:rest} \katoss{Plot of the estimated densities of the recovered mixture (\ref{eq:kj_mix_density}) (bold) and their components (dashed).}}
\end{figure}

\begin{table}
\caption{\label{tab:rest} \kotan{\katot{Maximum likelihood estimates \chris{for} the mixture with the recovered original parametrization (\ref{eq:kj_mix_density}).}}}
\centering
\fbox{
\begin{tabular}{*{6}{c}}
$k$ & $\hat{\mu}_k$ & $\hat{\gamma}_k$ & $\hat{\rho}_k$ & $\hat{\lambda}_k$ & $\hat{\pi}_k$ \\
\hline
1 & 2.7572 & 0.3751 & 0.7267 & 5.3136 & 0.4845 \\
2 & 4.0107 & 0.4855 & 0.1970 & 1.1895 & 0.5155 \\
\end{tabular}}
\end{table}

\kota{
Here, the parameters are estimated in the form of the mixture of two Kato--Jones distributions and the uniform distribution (Table \ref{tab:em}(b) and equation (4)).
Actually, we can regard this mixture of three components as the estimated \katot{model}.
Nonetheless, we are also able to reparameterize and recover the estimates of \katot{the} original parameters \katot{(\ref{eq:kj_mix_density})} as discussed in Section 4.1.
As we initially aim to estimate the mixture of two components, we employ the \chriss{latter} in the following.}
% Note that the model with uniform distribution (Table \ref{tab:emm}(b)) is one of the choices.
% It can be reparameterized and interpreted as well.
% In this study, the recovered model is employed from the viewpoint of interpretability.
The \katot{maximum likelihood estimates \chris{obtained via the} EM algorithm} (Fig.\ \ref{fig:em} and Table \ref{tab:em}) are recovered \katot{to} \katot{$\hat{\mu}_k, \hat{\gamma}_k, \hat{\rho}_k, \hat{\lambda}_k$ and ${\hat{\pi}}_k$} \kotann{as shown in Fig.\ \ref{fig:rest} and Table \ref{tab:rest}}.
The \katot{shapes of the two estimated components with $k=1,2$ in Table \ref{tab:em}  slightly change} by adding the restored uniform \katot{component with $k=3$.}

\kotar{In addition, $\hat{\mu}_k, \hat{\gamma}_k, \hat{\rho}_k, \hat{\lambda}_k$ are reparameterized into $\hat{\mu}_k, \hat{\gamma}_k, \hat{\bar{\alpha}}_{2k}, \hat{\bar{\beta}}_{2k}$ for interpretation because
$\hat{\bar{\alpha}}_{2k}$ and $\hat{\bar{\beta}}_{2k}$ can be interpreted as the circular skewness and kurtosis of \citet{bat} as mentioned in Section 3.2.}
The recovered parameters are reparameterized as shown in Table \ref{tab:repa}.
\kotar{The discussion of the parameter values to follow is carried out only for the parameters in Table \ref{tab:repa} because of the interpretability for $\hat{\bar{\alpha}}_{2k}$ and $\hat{\bar{\beta}}_{2k}$.}

\begin{table}
\caption{\label{tab:repa} \kotan{\katot{Maximum likelihood estimates \chris{for} the reparametrized mixture (\ref{eq:repa_density}).}}}
\centering
\fbox{
\begin{tabular}{*{6}{c}}
$k$ & $\hat{\mu}_k$ & $\hat{\gamma}_k$ & $\hat{\bar{\alpha}}_{2k}$ & $\hat{\bar{\beta}}_{2k}$ & $\hat{\pi}_k$ \\
\hline
1 & 2.7572 & 0.3751 & 0.1542 & -0.2248 & 0.4845 \\
2 & 4.0107 & 0.4855 & 0.0356 & 0.0888 & 0.5155 \\
\end{tabular}}
\end{table}

\subsection{\kota{Discussion}}

Here, we discuss the interpretation of the estimated models with the recovered parametrizations (\ref{eq:repa_density}) given in Table \ref{tab:repa}.
As shown in Fig.\ \ref{fig:rest}, the densities of the mixture fit fairly well to the histogram for the entire day, such as the sharp peak in the morning and the gentle peak in the evening.
The numerically calculated morning mode and evening mode of \kotar{the whole distribution of }the mixture are 7:32 and 15:56, respectively.

The two components divide the mixture into 48.5$\%$ and 51.5$\%$ as denoted by each value of $\pi_k$.
Each peak roughly represents the morning and evening traffic.
The peak of the component with $k=1$ is sharp and negatively skewed.
Its mean, $\hat{\mu}_1$, is around 10:32, and its mode is around 7:28 \citep[Supplementary Material]{kat15}.
The mode of the component with $k=1$ is close to the morning mode of the mixture because the component with $k=1$ is dominant around the morning.
The shape is characterized by large $\hat{\bar{\alpha}}_{21}$ and negatively large $\hat{\bar{\beta}}_{21}$.
On the other hand, the peak of the component with $k=2$ is gentle and positively skewed.
Its mean, $\hat{\mu}_2$, is around 15:19, and its mode is around 16:37.
The mode of the component with $k=2$ and the evening mode of the mixture are apart because the tail of the component with $k=1$ is contained even in the evening part of the mixture.
The shape is characterized by small $\hat{\bar{\alpha}}_{22}$ and positively small $\hat{\bar{\beta}}_{22}$.

The peak period of the component with $k=1$ is from 7:00 to 8:00, and its shape is remarkably sharp and quick-rising, which is described by large $\hat{\bar{\alpha}}_{21}$ and negatively large $\hat{\bar{\beta}}_{21}$.
This is explained by the actual situation that many drivers would like to arrive in central Osaka around the same time, like 8:30 and 9:00, which is the typical start time for working in Japan.
On the other hand, the peak period of the component with $k=2$ is from 15:00 to 19:00, and its shape is gradual and positively skewed, which is described by small $\hat{\bar{\alpha}}_{22}$ and positively small $\hat{\bar{\beta}}_{22}$.
This may be because drivers do not have to leave their office at the same time, which is different from morning hours.
Some of them may decide to make their departure time earlier or later than peak hour to avoid traffic congestion.
In addition, the positively skewed shape may imply that they also avoid returning home too late.

These interpretations are made by focusing on the shape around the mode of each peak.
It is more difficult to understand the traffic flow volume around the tails of both components.
For example, the first component still has a moderate probability at around 16:00, which cannot be included in the morning rush hour.
One possible explanation is that the first component represents the outward trip, including morning commuting, commercial vehicles, etc.
For example, many commercial vehicles travel from Kobe to Osaka during the daytime.
In addition, some people make their first trip in the evening because they work at night or go shopping.
People who work at night are more likely to use their vehicles for commuting than those who work in the daytime, as public transport does not operate after midnight.
This could explain the first component having a larger probability than the second component after around 22:00.
Similarly, the second component represents the day's return trips, including evening commuting.
As round and migration trips during the daytime, for example, may be included in the second component, it already has a moderate value even in the morning hours.
Note that outward and return trips are observed only once because all recorded traffics are in the same direction.

Finally, we compare our results with some phenomena typically mentioned in the traffic engineering field.
First, the difference in the value of $\hat{\bar{\alpha}}_{2}$ between the two components is consistent with the concept of departure time choice in traffic engineering \citep{nol95}.
In general, drivers choose their departure time by estimating backward from the time they want to arrive.
\cite{ale15} indicate that many drivers depart within short periods in the morning (i.e., traffic is concentrated), while drivers depart within longer periods in the evening and later.
Second, 30$\%$ of Hanshin Expressway users are for commercial purposes \citep[(in Japanese)]{han}.
Assuming that people make outward commercial trips after they start working, the tail of the first components around noon is likely to contain commercial vehicles at a certain level.
Therefore, the interpretation mentioned above that the first component is for the outward trips of the day and the second one is for the return trips is reasonable to some extent.
Note that the share of large vehicles on this road is 13.6$\%$ in total and less than 30$\%$ even in the midnight period, although one might expect that the freight traffic is the main component after midnight (see Appendix B Supplementary Material).
Therefore, no peak for freight traffic in the midnight period is found in the histograms, nor are such peaks found in the estimated model.
Since these discussions are only derived from traffic volume data, it is necessary to further develop the interpretations by combining trip purpose data and origin and destination data.
Nonetheless, it would be a contribution of this paper that the proposed model that only uses traffic volume information is consistent with various concepts used in traffic engineering.

\section{Comparison with other models} \label{sec:comparison}

\subsection{Mixtures of \kotar{other distributions}} \label{sec:comparison_other}

\kotar{We compare other \kota{multimodal} distributions with our proposed model in terms of \katot{practicability and fit}.
The distributions considered here are the mixture of von Mises distributions (MovM), the mixture of wrapped Cauchy distributions (MowC), the mixture of sine-skewed von Mises distributions (MossvM) and the mixture of sine-skewed wrapped Cauchy distributions (MosswC) whose densities are given by}
\begin{eqnarray*}
f_{MovM}(\theta) &=&  \sum_{k=1}^m \frac{\pi_k}{2\pi I_{0}(\kappa_k)} \exp(\kappa_{k}\cos(\theta-\mu_k)),\\
f_{MowC}(\theta) &=&  \sum_{k=1}^m \frac{\pi_k}{2\pi} \frac{1-\rho_k^2}{1+\rho_k^2-2\rho_k\cos{(\theta-\mu_k)}},\\
f_{MossvM}(\theta) &=&  \sum_{k=1}^m \frac{\pi_k(1+\lambda_k\sin(\theta-\mu_k))}{2\pi I_{0}(\kappa_k)} \exp(\kappa_{k}\cos(\theta-\mu_k)),\\
f_{MosswC}(\theta) &=&  \sum_{k=1}^m \frac{\pi_k}{2\pi} \frac{1-\rho_k^2}{1+\rho_k^2-2\rho_k\cos{(\theta-\mu_k)}}(1+\lambda_k\sin(\theta-\mu_k)),\\
\end{eqnarray*}
respectively, where $\mu_k \in [0, 2\pi)$ is the mean direction, \kotar{$\kappa_k \in [0, \infty)$ and $\rho_k \in [0, 1]$ are concentration parameters, $\lambda_k \in [-1, 1]$ is a skewing parameter}, $I_{0}(\cdot)$ is the modified Bessel function of the first kind and order zero\chris{,} $m \in \mathbb{N}$ is the number of components of the mixture and $0 < \pi_1,\ldots,\pi_m < 1$ are the weights of the components satisfying $\sum_{k=1}^m \pi_k =1 $.
\kotar{The number of free parameters for MovM and MowC is $3m-1$, and that for MossvM and MosswC is $4m-1$.
These distributions are popular mixtures discussed, e.g., in \citet{wal00,moo03,ban05,mul20,miy20}.
Note that the mixture of sine-skewed wrapped Cauchy is a submodel of the proposed model \cite[Supplementary Material]{kat15}.}

\kotar{The 50-fold cross-validation is conducted to compare the performance of the proposed model with that of these four existing models.
The cross-validated log-likelihood is employed as the criterion for comparison \citep{smy00}.
The number of components is set to two, as in our proposed model.}

\begin{table}
\caption{\label{tab:cross1} \kotar{The average value of cross-validated log-likelihood function in 50-fold cross validation for MovM, MowC, MossvM, MosswC and proposed model.}}
\centering
\fbox{
\begin{tabular}{*{4}{c}}
model & log-likelihood function\\
\hline
proposed & -36818.5\\
MovM & -37059.4\\
MowC & -37731.7\\
MossvM & -37026.3\\
MosswC & -36855.3\\
\end{tabular}}
\end{table}

\begin{figure}[t]
    \begin{center}
			\includegraphics[clip, keepaspectratio, width=0.75\hsize]{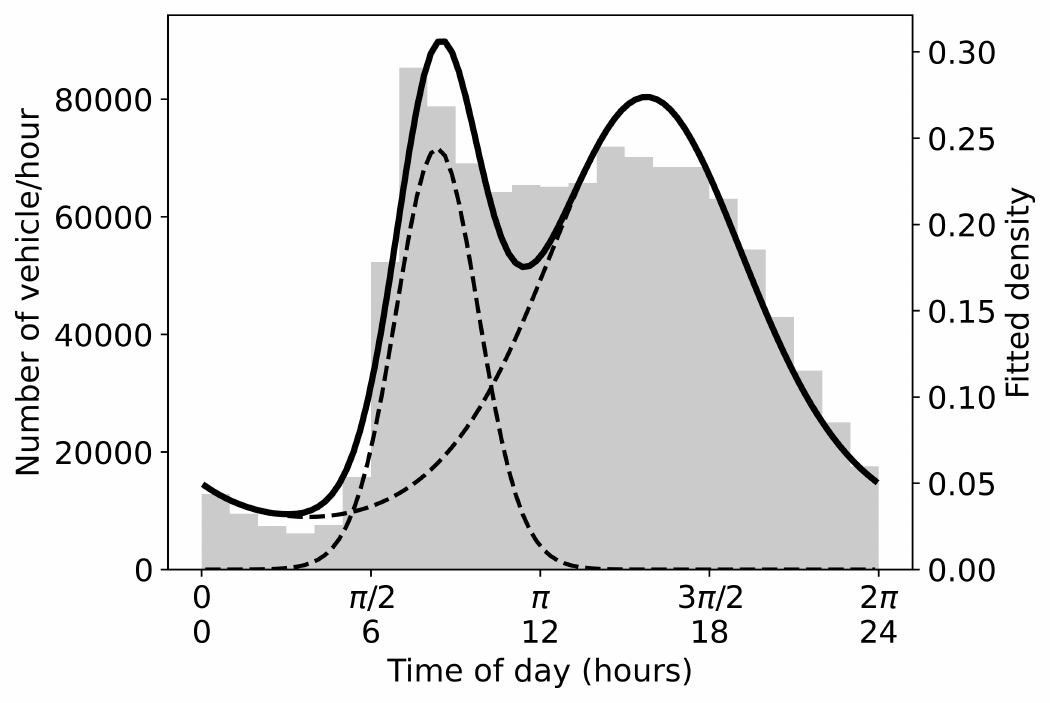}
    \end{center}
\caption{\label{fig:vm} Plot of maximum likelihood fit of the density of MovM with two components.}
\end{figure}

\kotar{
Table \ref{tab:cross1} shows each model's average value of the cross-validated log-likelihood function.
The predictive performance of the proposed model is superior to the other models because the value for the proposed model is the largest.
\kotarr{
As an example of a plot of the estimated densities, Fig.~\ref{fig:vm} shows the maximum likelihood fits of the densities of MovM, the most well-known mixture of circular distributions, with two components.
The goodness of fit looks significantly worse than in the proposed model shown in Fig.~\ref{fig:rest}.
Particularly, the distorted peak around $\pi/2$ and the gradual peak around $3\pi/2$ are not represented.}
In contrast to the proposed model that allows varying kurtosis and skewness values, MovM and MowC do not allow both, and MossvM and MosswC do not allow varying kurtosis.
In terms of the number of parameters, MovM and MowC have $m$ fewer than the proposed model, and MossvM and MosswC have the same.
The proposed model is superior to the other models in terms of predictive performance because the proposed model can vary its shape flexibly with a small number of parameters.}
\kotarr{In addition to the cross-validation, a comparison of the proposed model with other models using simulation studies is provided in Appendix C in Supplementary Material.}

\subsection{Our \katot{mixture with more than two components}}%m=8くらいまで計算しましたがどうしますか？
%We assume two components in Section 5 because two typical peaks can be recognized in the histogram.
\katot{We estimate} the mixed Kato--Jones distributions \katot{(\ref{eq:repa_density})} whose number of components $m$ \katot{is} more than two.
\kota{In this \katot{subsection}, models \katot{with $m \leq 6$ are discussed} because of stability of estimation; estimation for the model \katot{with  $m > 6$} \chris{was} carried out but the \chris{results were} unstable.}
\chris{Recall} that the histogram shows two peaks that can be interpreted as shown above and the assumption of more than two peaks is not necessarily insightful from the transport engineering viewpoint.
As such, the main purpose \chris{of} this subsection is not to find the most accurate number of components but \chris{to find out} what \chris{shapes} of distributions will be \chris{obtained}.
%However, there are not only two types traffic; morning traffic and evening traffic, but also another kind of traffic such as commercial traffic, long-distance lorry and journey.
%Therefore, other models whose number of components is more than 2 are estimated.

\kotar{The 50-fold cross-validations for the proposed model, whose number of components $m$ is from 2 to 6, are conducted.}
\kotan{The conditions for the estimation are the same as those in Section 5.}

\begin{table}
\caption{\label{tab:cross2} \kotar{The average value of cross-validated log-likelihood function in 50-fold cross validation for proposed model whose number of components $m$ from 2 to 6}}
\centering
\fbox{
\begin{tabular}{*{4}{c}}
$m$ & log-likelihood function\\
\hline
2 & -36818.5\\
3 & -36818.7\\
4 & -36808.6\\
5 & -36817.6\\
6 & -36820.2
\end{tabular}}
\end{table}

\begin{figure}[t]
    \begin{center}
			\includegraphics[clip, keepaspectratio, width=0.75\hsize]{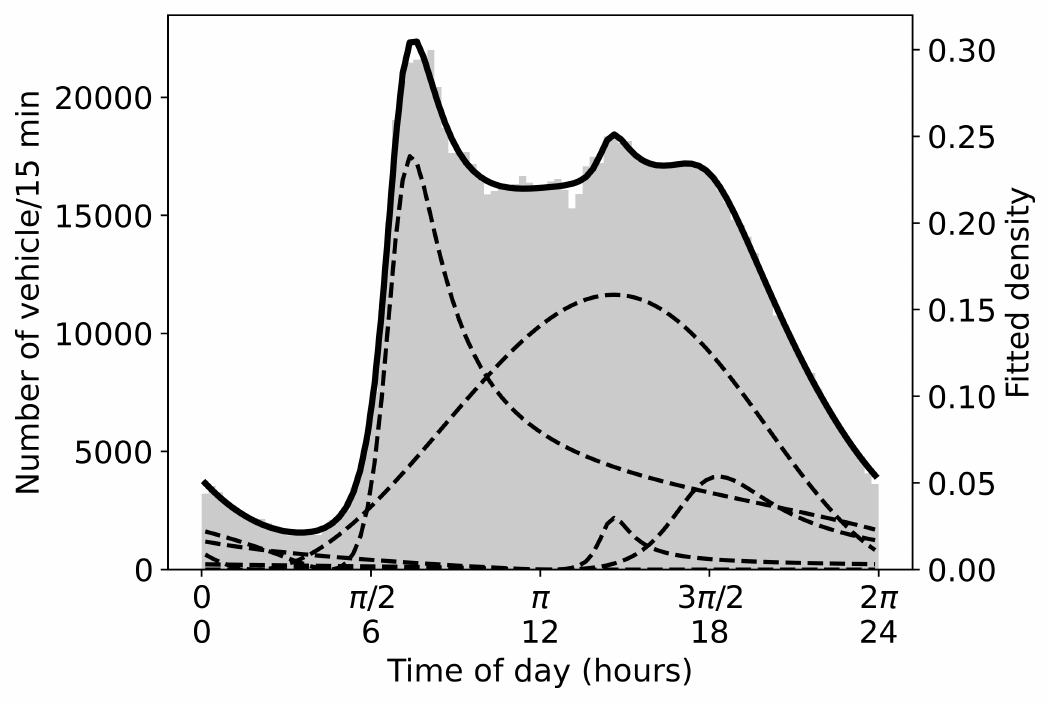}
    \end{center}
\caption{\label{fig:m4} \katoss{Plot of maximum likelihood fits of the densities with the numbers of components $m=4$.}}
\end{figure}

\kotar{
Table \ref{tab:cross2} shows the average value of the cross-validated log-likelihood function for each number of components.
The model whose number of components $m=4$ is the best because the value for the proposed model is the largest.
The plot and value of the parameter for the estimated model whose number of components $m=4$ are shown in Fig.\ \ref{fig:m4} and Table \ref{tab:m4}.
Note that the number of bins in Fig.\ \ref{fig:m4} is 96 (i.e., the width of the bin equals 15 minutes) to visualise the small peak around 15:00 described by the component with $k=3$.}

\kotar{
In the results of $m=4$, two prominent components with $k=1, 2$ have a large value of $\hat{\pi}_k$.
As well as the results of $m=2$, the components with $k=1,2$ represent the peak around $\pi/2$ and $3\pi/2$, respectively.
However, the component with $k=2$ for $m=4$ is not skewed compared to the component with $k=2$ for $m=2$ shown in Fig.\ \ref{fig:rest}.
Not only the component with $k=2$ but also that with $k=4$ in the results of $m=4$ represent the shape between $3\pi/2$ and $2\pi$ that is represented by only one component with $k=2$ in the results of $m=2$.
In addition, the component \katot{with $k=3$} has \chris{a} small value of \katot{$\hat{\pi}_3$} and represents the small peak around 16:00 seen in the histogram.}

\kotar{
The components with $k=3,4$ may not be meaningful from the transport engineering viewpoint;
they only represent the trivial peak along \chris{with} the improvement of the value of \chris{the} log-likelihood function.
The interpretation of the parameters for these smaller components is difficult with only the traffic volume data.
Some additional information, such as each vehicle's origin, destination, and trip purpose, may help the interpretation of the parameters.}
%Further consideration of determining the appropriate number of components may be necessary.
%The values of AIC and BIC suggest a larger number of components.
%The small components seen in these cases might be generated by some unknown factors and phenomena.
%Nonetheless, from the viewpoints of interpretability of the result and feasibility of the calculation, we regard $m=2$ as the most reasonable numbers of components in this study.

\begin{table}
\caption{\label{tab:m4} Maximum likelihood estimates of the parameters of Kato--Jones mixtures (\ref{eq:repa_density}) whose numbers of components $m=4$.}
\centering
\fbox{
\begin{tabular}{*{6}{c}}
$k$ & $\hat{\mu}_k$ & $\hat{\gamma}_k$ & $\hat{\bar{\alpha}}_{2k}$ & $\hat{\bar{\beta}}_{2k}$ & $\hat{\pi}_k$ \\
\hline
 1 & 2.6665 & 0.4376 & 0.2128 & -0.2452 & 0.3855 \\
 2 & 3.7380 & 0.5065 & 0.0076 & 0.0225 & 0.4907 \\
 3 & 4.3642 & 0.4887 & 0.3402 & -0.2284 & 0.0285 \\
 4 & 5.2889 & 0.6037 & 0.2628 & -0.2166 & 0.0952 \\
\end{tabular}}
\end{table}

\section{Conclusion} \label{sec:conclusion}

In this paper, we estimated the probability distribution of the variation of traffic volume within an average weekday.
We employed the mixture of Kato--Jones distributions, which 
%is one of the circular distributions and 
can provide a wide range of skewness and kurtosis \katot{for each component}.
%for the variation of traffic volume within an average weekday with using the raw data of traffic counter.
To estimate the parameters of these distributions, we developed the modified method of moments.
\kota{
This method \chris{ensures} that the estimated parameters 
% One of the advantages of the method is that the estimate 
always belong to the parameter space.
}
\katos{Then the maximum likelihood method for the proposed mixture was established using the EM algorithm}.

The proposed method was applied to \chris{some} traffic counter data \chris{from} Japan. 
% In the result of the number of components $m=2$, 
As a result, the data were classified into two components, which were interpreted as the morning and evening traffic.
%by the proposed method.
Some reasonable explanations for these components are suggested\chris{,} that the difference in \chris{the} shapes \chris{of the components} might be caused by the different \chris{behaviours underlying} the commuting trips between morning and evening hours.
\cyan{In addition, the modified method of moments was \katoss{seen} to allow for fast calculation and provided \chris{a} reasonable initial value \chris{for} maximum likelihood estimation.}
\katos{With this initial value, the EM algorithm provided reasonable maximum likelihood \chris{estimates}.}
% the method works quickly and the estimate from the method is a useful 

% The first distribution is negatively skewed peak whose mode is 7:28.
% This result is consistent with the actual situation that many drivers would like to arrive central Osaka in just around the same time as 8:30 and 9:00.
% On the other hand, the second distribution is gentle and positively skewed peak whose mode is 16:37.
% This result shows that evening traffic does not concentrate compared with morning one.
% This may be because drivers do not have to leave their office at the same timing.
% We estimate the model with another number of components from $m=3$ to $m=8$.
% The value of $AIC$ and $BIC$ get better as $m$ increases.
% On the other hand, the larger $m$ needs the longer computation time.
% Moreover, the interpretation of the all components is difficult.
%From the viewpoints of interpretability of the result and the feasibility of the calculation, we regard that $m=2$ is one of the reasonable number of components in this study.
%According to \katot{information criteria} such as $AIC$ and $BIC$, \chris{a} larger number of components than two resulted in better \chris{estimation}.
\kotar{According to the cross-validated log-likelihood function, the model with the number of components $m=4$ is the best.
However, the interpretation of the trivial components in the model is difficult thus far.
Further consideration of missing factors that generate third and later components may be necessary in future work.
Additional information, such as each vehicle's origin, destination, and trip purpose, would be useful.
In addition, we assume the IID property for all traffic data in the modeling, and we regard this assumption as acceptable, as mentioned in the supplementary Material.
However, modeling the dependent characteristics of the arrival interval, if it exists, is one of the principal and important issues to be dealt with in the future.}

% However, if we have another data such as the data of the origin of each vehicle or the purpose of their trip, the more interpretation of another number of components can be improved.

\section*{Acknowledgement}
\kotarr{We thank an associate editor and two reviewers for their instructive and insightful comments on an earlier version of this paper.}
\katot{The authors are grateful to Hironori Fujisawa for constructive comments on this work.}
The data used in this study were provided by \chris{the} Hanshin Expressway Co., Ltd.
This research was partially supported by JSPS Kakenhi \katos{18K13846, 20H02267 and 20K03759}.

\section*{Supplementary Material}
Supplementary Material is available online at \textit{Journal of the Royal Statistical Society: Series C}.


\begin{thebibliography}{7}

	%\expandafter\ifx\csname natexlab\endcsname\relax\def\natexlab#1{#1}\fi

	\bibitem[\protect\citeauthoryear{Alexander \textit{et al.}}{2015}]{ale15}
	Alexander, L., Jiang, S., Murga, M. and González, M.C. (2015).
	Origin–destination trips by purpose and time of day inferred from mobile phone data.
	\textit{Transportation research part c: emerging technologies}, \textbf{58}, 240--250.
 
	\bibitem[\protect\citeauthoryear{Anacleto \textit{et al.}}{2013}]{ana13}
	Anacleto, O., Queen, C. and Albers, C. J. (2013).
	Multivariate forecasting of road traffic flows in the presence of heteroscedasticity and measurement errors.
	\textit{Journal of the Royal Statistical Society: Series C (Applied Statistics)}, \textbf{62}, 251--270.
	
	\kator{
	\bibitem[\protect\citeauthoryear{Banerjee \textit{et al.}}{2005}]{ban05}
	Banerjee, A., Dhillon, I.S., Ghosh, J. and Sra, S. (2005).
	Clustering on the unit hypersphere using von Mises--Fisher distributions.
	\textit{Journal of Machine Learning Research}, \textbf{6}, 1345--1382.
	}
	
	
\kato{
	\bibitem[\protect\citeauthoryear{Batschelet}{1981}]{bat}
	Batschelet, E. (1981).
	\textit{Circular Statistics in Biology.} 
	London: Academic Press.
	}

	\bibitem[\protect\citeauthoryear{Chen \textit{et al.}}{2010}]{che10}
	Chen, X., Li, L. and Zhang, Y. (2010).
	A Markov model for headway/spacing distribution of road traffic.
	\textit{IEEE Transactions on Intelligent Transportation Systems}, \textbf{11}, 773--785.
	
	\bibitem[\protect\citeauthoryear{Daganzo}{1997}]{dag97}
	Daganzo, C. F. (1997).
	\textit{Fundamentals of transportation and traffic operations.} 
	Pergamon Oxford.

	\bibitem[\protect\citeauthoryear{Davison and Hinkley}{1997}]{dav97}
	Davison, A. C. and Hinkley, D. V. (1997).
	\textit{Bootstrap methods and their application.} 
	Cambridge university press.
	
	\bibitem[\protect\citeauthoryear{Edie}{1963}]{edi63}
	Edie, L.C. (1963).
	\textit{Discussion of traffic stream measurements and definitions.}
	Port of New York Authority.
        \kotar{\bibitem[\protect\citeauthoryear{Hanshin Expressway}{2017}]{han}
	Hanshin Expressway (2017).
	\textit{CSR report on 2017.}
	viewed 25 May 2023 \url{https://www.hanshin-exp.co.jp/company/csr/files/report/csrreport_201707_2.pdf.}}
\katott{
\bibitem[\protect\citeauthoryear{Huber}{1981}]{hub}
	Huber, P.J. (1981).
	\textit{Robust Statistics}. New York: Wiley.
 }

	\bibitem[\protect\citeauthoryear{Jones \textit{et al.}}{2001}]{jon01}
	Jones, E., Oliphant, T., and Peterson, P. (2001).
	SciPy: Open source scientific tools for Python.
	(Available from http://www.scipy.org/)
	
	\bibitem[\protect\citeauthoryear{Kato and Jones}{2015}]{kat15}
	Kato, S. and Jones, M.C. (2015).
	A tractable and interpretable four-parameter family of unimodal distributions on the circle.
	\textit{Biometrika}, \textbf{102}, 181--190.
	
	\bibitem[\protect\citeauthoryear{Kraft}{1988}]{kra88}
	Kraft, D. (1988).
	A software package for sequential quadratic programming.
	(Available from http://www.opengrey.eu/item/display/10068/147127)

	\bibitem[\protect\citeauthoryear{Leduc}{2008}]{led08}
	Leduc, G. (2008).
	\textit{Road traffic data: Collection methods and applications.} 
	Working Papers on Energy, Transport and Climate Change.
	
	\bibitem[\protect\citeauthoryear{Ley \textit{et al.}}{2021}]{ley21}
	Ley, C., Babi\'c, S. and Craens, D. (2021).
	Flexible models for complex data with applications.
	\textit{Annual Review of Statistics and Its Application}, \textbf{8}, 369--391.

	\bibitem[\protect\citeauthoryear{Li and Chen}{2017}]{li17}
	Li, L. and Chen, X. M. (2017).
	Vehicle headway modeling and its inferences in macroscopic/microscopic traffic flow theory: A survey.
	\textit{Transportation Research Part C: Emerging Technologies}, \textbf{76}, 170--188.
	
	%\chris{\bibitem[\protect\citeauthoryear{Mardia and Jupp}{1999}]{mar99}
	%Mardia, K.V. and Jupp, P.E. (1999).
	%\textit{Directional Statistics}.
	%Chichester: Wiley.}
	
	\bibitem[\protect\citeauthoryear{McLachlan and Krishnan}{2007}]{mcl}
	McLachlan, G.J. and Krishnan, T. (2007).
	\textit{The EM Algorithm and Extensions}, 2nd ed.
	Hoboken: Wiley.
	
	\kator{
	\bibitem[\protect\citeauthoryear{Miyata \textit{et al.}}{2020}]{miy20}
	Miyata, Y., Shiohama, T. and Abe, T. (2020).
	Estimation of finite mixture models of skew-symmetric circular distributions.
	\textit{Metrika}, \textbf{83}, 895--922.
	}
	
	
	\kator{
	\bibitem[\protect\citeauthoryear{Mooney \textit{et al.}}{2003}]{moo03}
	Mooney, J.A., Helms, P.J. and Jolliffe, I.T. (2003).
	Fitting mixtures of von Mises distributions: a case study involving sudden infant death syndrome.
	\textit{Computational Statistics \& Data Analysis}, \textbf{41}, 505--513.
	}
	
	
	\kator{
	\bibitem[\protect\citeauthoryear{Mulder \textit{et al.}}{2020}]{mul20}
	Mulder, K., Jongsma, P. and Klugkist, I. (2020).
	Bayesian inference for mixtures of von Mises distributions using reversible jump MCMC sampler.
	\textit{Journal of Statistical Computation and Simulation}, \textbf{90}, 1539--1556.
	}

	\bibitem[\protect\citeauthoryear{Muñoz \textit{et al.}}{2003}]{mun03}
	Muñoz, L., Sun, X., Horowitz, R. and Alvarez, L. (2003).
	Traffic density estimation with the cell transmission model.
	\textit{In Proceedings of the 2003 American Control Conference}, \textbf{5}, 3750--3755.
	
	\bibitem[\protect\citeauthoryear{Na and Jang}{2011}]{na11}
	Na, J. H. and Jang, Y. M. (2011).
	Modeling on daily traffic volume of local state road using circular mixture distributions.
	\textit{The Korean Journal of Applied Statistics}, \textbf{24}, 547--557.
 
	%\kota{
	\bibitem[\protect\citeauthoryear{Nagasaki \textit{et al.}}{2022}]{nag22}
	Nagasaki, K., Kato, S., Nakanishi, W. and Jones, M. C. (2022).
	Traffic Count Data Analysis Using Mixtures of Kato--Jones Distributions on the Circle.
	\textit{arXiv preprint arXiv:2206.01355}.
 
	%\kota{
	\bibitem[\protect\citeauthoryear{Nagasaki \textit{et al.}}{2019}]{nag19}
	Nagasaki, K., Nakanishi, W. and Asakura, Y. (2019).
	Application of Rose Diagram to Road Network Analysis.
	\textit{In Proceedings of 24th International Conference of Hong Kong Society for Transportation Studies}, 169--176.
	%}

 	%\kota{
	\bibitem[\protect\citeauthoryear{Nagasaki \textit{et al.}}{2023}]{nag23}
	Nagasaki, K., Nakanishi, W. and Asakura, Y. (2023).
	Analysis of Traffic Volume by the Estimated Parameter of the Mixture of Circular Distribution.
	\textit{Journal of Japan Society of Civil Engineers, Ser. D3 (Infrastructure Planning and Management)}, \textbf{78}, I825--I831.


	\bibitem[\protect\citeauthoryear{Noland and Small}{1995}]{nol95}
	Noland, R.B. and Small, K.A. (1995).
	Travel-time uncertainty, departure time choice, and the cost of morning commutes.
	\textit{Transportation research record}, \textbf{1493}, 150--158.

	\katott{
 \bibitem[\protect\citeauthoryear{Pewsey and Garc\'ia-Portugu\'es}{2021}]{pew}
	Pewsey, A. and Garc\'ia-Portugu\'es, E. (2021).
	Recent advances in directional statistics. \textit{TEST}, \textbf{30}, 1--58.
 }
	
	\katorrr{
	\bibitem[\protect\citeauthoryear{Press}{1972}]{pre72}
	Press, S.J. (1972).
	Estimation in univariate and multivariate stable distributions.
	\textit{Journal of the American Statistical Association}, \textbf{67}, 842--846.
	}

 \katott{
	\bibitem[\protect\citeauthoryear{Redner and Walker}{1984}]{red84}
	Redner, R.A. and Walker, H.F. (1984).
 Mixture densities, maximum likelihood and the EM algorithm.
  \textit{SIAM Review}, \textbf{26}, 195--239.
	}

	\bibitem[\protect\citeauthoryear{Smyth}{2000}]{smy00}
	Smyth, P. (2000).
	Model selection for probabilistic clustering using cross-validated likelihood.
	\textit{Statistics and computing}, \textbf{10}, 63--72.
 
\bibitem[\protect\citeauthoryear{Transportation Research Board}{2010}]{hcm}
	Transportation Research Board (2010).
	\textit{Highway capacity manual.}
	National Research Council, Washington, DC.
	
	
	\kator{
	\bibitem[\protect\citeauthoryear{Wallace and Dowe}{2000}]{wal00}
	Wallace, C.S. and Dowe, D.L. (2003).
	MML clustering of multi-state, Poisson, von Mises circular and Gaussian distributions.
	\textit{Statistics and Computing}, \textbf{10}, 73--83.
	}
	
	\bibitem[\protect\citeauthoryear{Wang and Papageorgiou}{2005}]{wan05}
	Wang, Y. and Papageorgiou, M. (2005).
	Real-time freeway traffic state estimation based on extended Kalman filter: a general approach.
	\textit{Transportation Research Part B: Methodological}, \textbf{39}, 141--167.






	
	
	
	%\bibitem[\protect\citeauthoryear{Ley and Verdebout}{2017}]{ley17}
	%Ley, C. and Verdebout, T. (2017).
	%\textit{Modern Directional Statistics}.
	%Boca Raton: CRC Press.
	
	
\end{thebibliography}
\end{document}

\appendix

\section*{Appendix A.\quad Assumption of IID arrival}
The IID assumption of the vehicles arrival in this paper is acceptable because of the following reasons. 
First, if there is no congestion and all vehicles run at their desirable speed, the traffic flow is usually regarded as independent \citep{hcm}.
% HCM,7-3,SPEED 
The majority of the applied data belong to this situation. 
Next, even though there is a speed decline, almost all ``congestions'' in this dataset are not serious ones; 
all vehicles run at at least 20 or 30 [km/h] and no complete stop is observed. 
In addition, they have enough gap between the following and successive vehicles such as 5 to 10 [m]. 
In the transportation engineering field, this is called ``stagnation'' and distinguished from congestion.
It should be noted that each driver can somehow choose their speed in the stagnation \citep{brack09}. 
This directly means that the arrival timing is mainly decided by some randomness caused by each vehicle and driver. 

As for the identicality of the distribution, it is said that typical variation patterns of the vehicle arrival are related to the type of road and the day of the week \citep{hcm}.
In other words, the variation of vehicle type is weak compared to the within-day variation, which is dominant.
As a result, we assume that the vehicle arrivals are IID as a preliminary and first step modelling.
Of course, modelling the dependent characteristics of the arrival interval, if it exists, is one of the most important issues to be dealt with in the future.

\section*{Appendix B.\quad Share of trucks}

\begin{figure}[t]
\centering
\includegraphics[clip,keepaspectratio,width=0.75\textwidth]{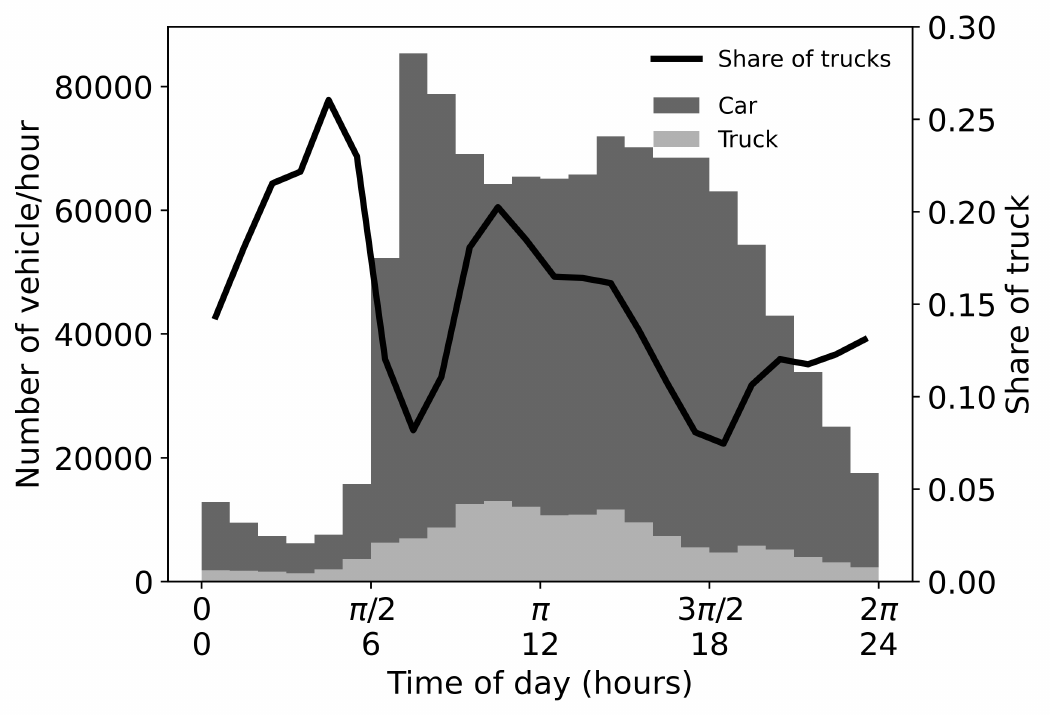}
\caption{The share of trucks to total traffic by hour (solid line).}\label{fig:share}
\end{figure}

\kotar{Fig.\ \ref{fig:share} shows the share of trucks to total traffic by the hour.
The total share of trucks is 13.6$\%$.
The share of trucks declined in the peak hours in the morning or evening.
On the other hand, it is increased around midnight and noon.
This may be because truck drivers would like to avoid the congestion.
However, the maximum share is smaller than 30 $\%$ around 4 a.m..
Therefore, the traffic of trucks is not considered dominant overall.}

\section*{Appendix C.\quad Simulation study for comparison to other mixtures}

We compare other multimodal distributions with our proposed model in terms of fit by simulation study.
The compared distributions are the mixture of von Mises distributions (MovM), the mixture of wrapped Cauchy distributions (MowC), the mixture of sine-skewed von Mises distributions (MossvM) and the mixture of sine-skewed wrapped Cauchy distributions (MosswC); see the Section 7.1. in the main article.

To evaluate the estimation performance as a mixture distribution, the values of the log-likelihood function of each mixture for the samples generated by different value of the weights of the components $\pi_k$ are compared.
First, random samples of sizes $n=100, 500, 1000, $ and $5000$ were generated from the mixture of Kato--Jones distributions whose parameters were $\mu_1=\pi/2, \rho_1=0.2, \lambda_1=-\pi/2, \mu_2=3\pi/2, \rho_2=0.7, \lambda_2=\pi/3$.
The set of $\{\pi_1',\pi_2'\}$ was $\{0.2,0.8\}$, $\{0.5,0.5\}$, $\{0.3,0.5\}$, $\{0.4,0.4\}$, $\{0.2,0.4\}$, and $\{0.3,0.3\}$.
For each sample size and the set of $\{\pi_1',\pi_2'\}$, $r=1000$ samples were generated.
Maximum likelihood estimation was performed on these samples for the aforementioned mixtures, and the value of log-likelihoods were compared.

\begin{figure}[t]
\subfigure[The sample size $n=100$.]{
	\begin{minipage}{0.48\hsize}
		\begin{center}
			\includegraphics[clip, keepaspectratio, width=\hsize]{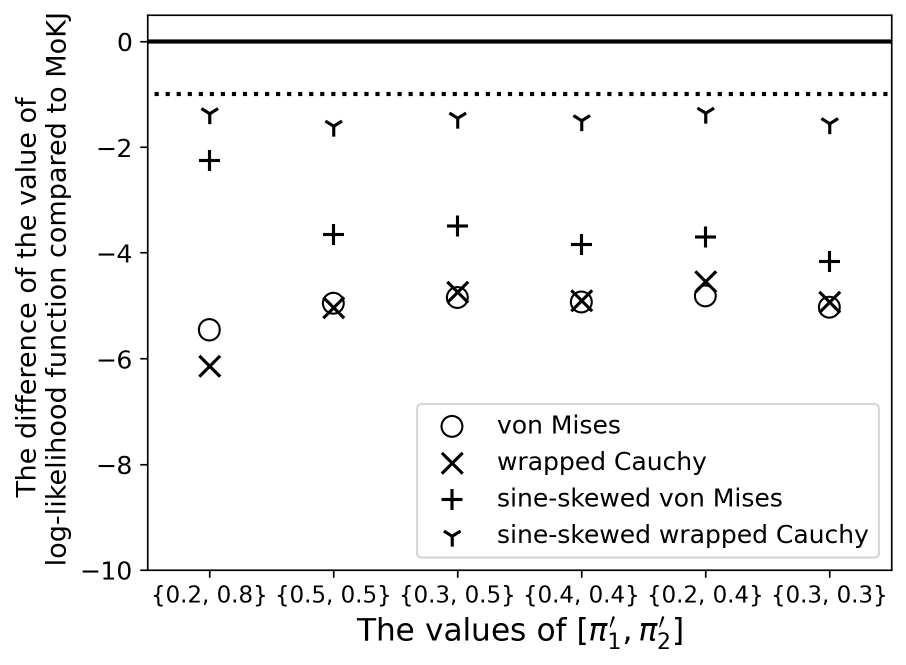}
		\end{center}
    \end{minipage}}
\subfigure[The sample size $n=500$.]{
	\begin{minipage}{0.48\hsize}
		\begin{center}
			\includegraphics[clip, keepaspectratio, width=\hsize]{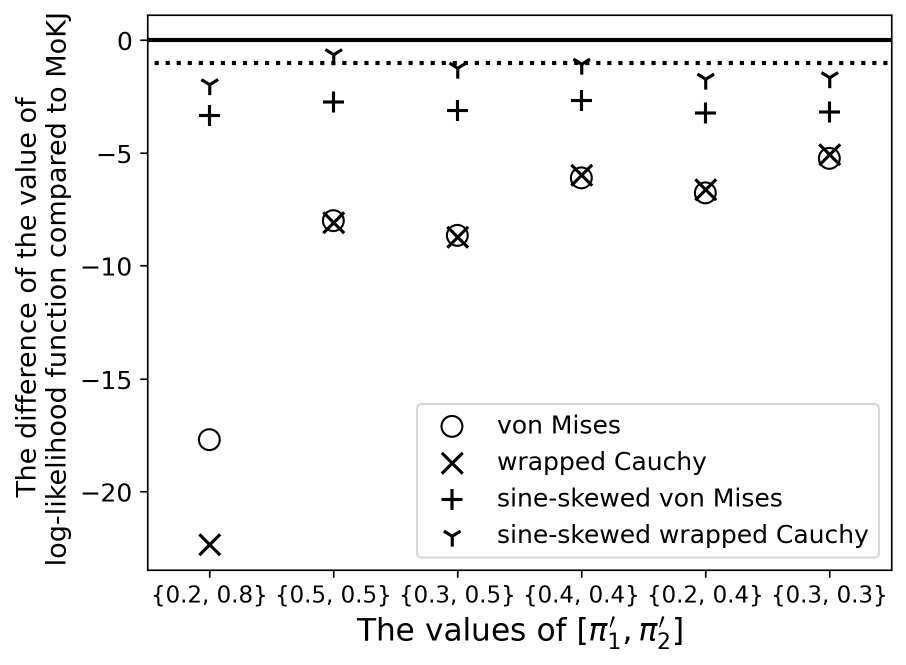}
		\end{center}
    \end{minipage}}\\
\subfigure[The sample size $n=1000$.]{
	\begin{minipage}{0.48\hsize}
		\begin{center}
			\includegraphics[clip, keepaspectratio, width=\hsize]{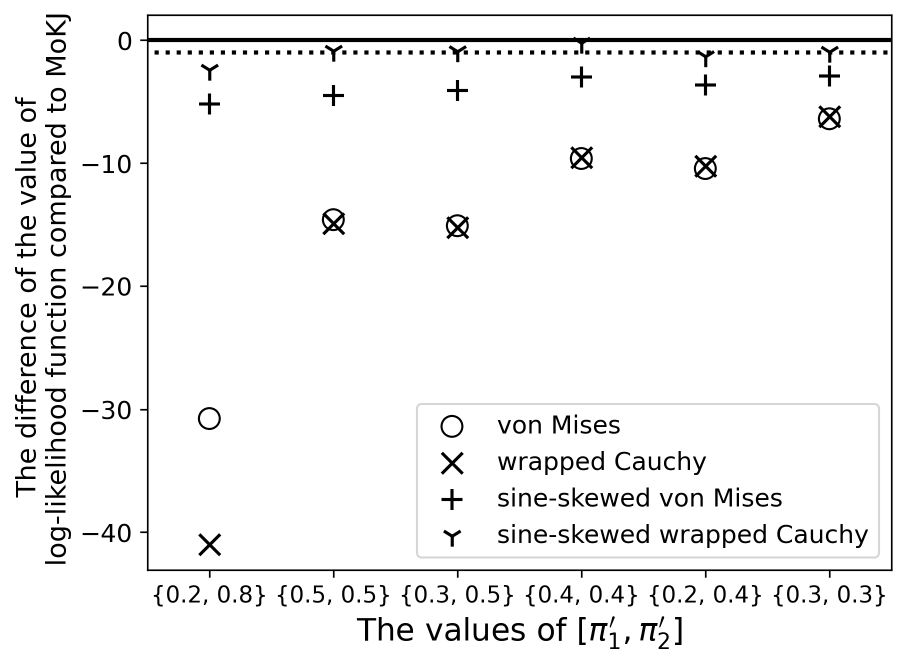}
		\end{center}
    \end{minipage}}
\subfigure[The sample size $n=5000$.]{
	\begin{minipage}{0.48\hsize}
		\begin{center}
			\includegraphics[clip, keepaspectratio, width=\hsize]{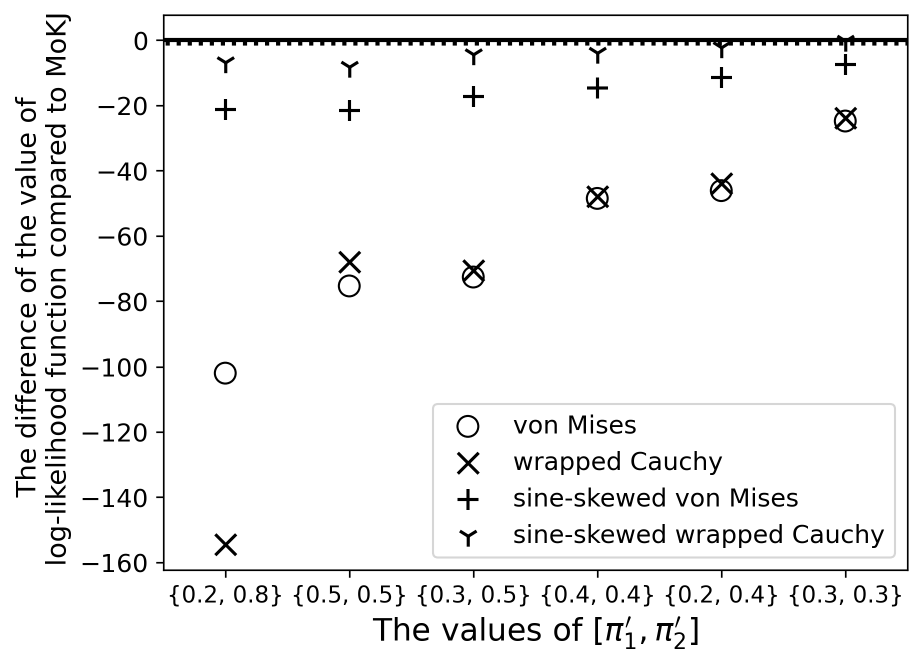}
		\end{center}
    \end{minipage}}
\caption{Panels from (a) to (d) show the difference of the value of log-likelihood function compared to MoKJ for each sample size and the set of $\{\pi_1',\pi_2'\}$.}\label{fig:difllhsim}
\end{figure}

Fig.\ \ref{fig:difllhsim} shows the difference of the value of the log-likelihood function compared to MoKJ for each sample size and the set of $\{\pi_1',\pi_2'\}$.
In the case of MosswC, the plots above the dotted line where the difference equals -1 are better models than MoKJ in terms of AIC.

Since MowC and MosswC are submodels of MoKJ and MovM and MossvM cannot describe skewness, the log-likelihood function of MoKJ was the best value for all samples.
In addition, some results show that MosswC is superior to MoKJ in terms of AIC.
For a small sample size, such as $n=100$, MoKJ is superior for all sets of $\pi_k'$, whereas when the number of samples is larger, MosswC is superior in some cases depending on the set of $\pi_k'$.
However, MoKJ is also suggested to be a better model in many cases.
MoKJ outperforms all other models, namely MovM, MowC, and MossVM, in terms of AIC across all sample sizes and values of $(\pi_1',\pi_2')$.
For both MovM and MowC, the difference in log-likelihood values relative to MoKJ tends to increase with $\pi_1' + \pi_2'$.
Note that while this trend was also observed in other cases with different sample sizes and parameter settings, the estimation of alternative models did not converge stably for some sample sizes and parameter settings. 

\section*{Appendix D.\quad Proofs}

\katott{
\subsection*{D.1.\quad Proof of Theorem \ref{thm:skewness}} \label{sec:proof_skewness}

The $p$th trigonometric moment of a random variable $\Theta$ following the Kato--Jones distribution (\ref{eq:kj_density}) is $\phi_{\rm KJ}(p) \equiv E(e^{i p \Theta}) = \gamma \,(\rho e^{i \lambda})^{-1}  \left\{ \rho e^{i (\mu + \lambda)} 
\right\}^p $ \citep[equation (2)]{kat15}.
For convenience, we reparametrize the Kato--Jones distribution (\ref{eq:kj_density}) via $\alpha = -\lambda$, $\beta=\gamma/\rho$ and $\eta=\mu+\lambda$.
Then we have $\phi_{\rm KJ} (p) =
\beta e^{i \alpha} \left( \rho e^{i \eta} \right)^p$.

%Then the , using the Fourier series expansion, the density $h_j$ can be expressed as
%$$
%h_j (\theta) = \frac{1}{2\pi} \sum_{q=-\infty}^{\infty} \xi_{j,q} e^{-i q (\theta-\mu_j) } ,
%$$
%where $-\pi \leq \mu_j < \pi$ and $|\xi_{j,q}| \leq 1 $ \citep[see][equation (3.4.25)]{mar99}.
The $p$th trigonometric moment of a random variable having the density $h_j$ is given by $\phi_{h_j} (p) = \int_{-\pi}^{\pi} e^{i p \theta} h_j(\theta) d\theta \ = e^{i p \mu_j} \int_{-\pi}^{\pi} e^{i p (\theta-\mu_j)} h_j (\theta) d\theta = \xi_{j,p} e^{i p \mu_j}$, where $\xi_{j,p} = E [\cos\{p (\theta-\mu_j)\}]$.
It follows from the assumption that $\xi_{j,p} > 0$ for any $p \in \mathbb{N}$.
We show that, if
$$
\phi_{\rm KJ}(p) = \pi_1 \phi_{h_j} (p) + \cdots + \pi_m \phi_{h_m}(p), \label{eq:proof_theorem}
$$
then $\alpha =0$.
The above equation is equivalent to
$$
\beta \rho^p e^{i \alpha} = \pi_1 \xi_{1,p} e^{i p (\mu_1 - \eta)} + \cdots + \pi_m \xi_{m,p} e^{i p (\mu_m - \eta)}.
$$
Taking arguments of both sides of this equation, we have
\begin{equation}
\alpha = \arg \left\{  \pi_1 \xi_{1,p} e^{i p (\mu_1-\eta)} + \cdots + \pi_m \xi_{m,p} e^{i p (\mu_m-\eta)} \right\}, \quad p \in \mathbb{N}. \label{eq:alpha}
\end{equation}
%It follows that
%$$
%\arg \left\{ \pi_1 \xi_{1,p} e^{i \{ p(\mu_1-\eta) - \alpha \}} + \pi_2 \xi_{2,p} e^{i \{ p(\mu_2-\eta) - \alpha \}} \right\} = 0.
%$$
From the assumption that there exist $a_k \in \mathbb{Z}$ and $b_k \in \mathbb{N}$ such that $\mu_k-\eta = a_k/b_k$, the equation (\ref{eq:alpha}) with $p=2 b_1 \cdots b_m$ reduces to
$$
\alpha = \arg \{ \pi_1 \xi_{1,p} + \cdots + \pi_m \xi_{m,p} \} = 0.
$$
Thus it follows that $\alpha= - \lambda = 0$, which corresponds to a symmetric case of the Kato--Jones distribution. \hfill $\Box$
}

\katott{

\subsection*{D.2.\quad Proof of Theorem \ref{thm:skewness_wc}} \label{sec:proof_skewness_wc}

As in the proof of Theorem \ref{thm:skewness}, the $p$th trigonometric moment of a random variable having the Kato--Jones distribution (\ref{eq:kj_density}) is $ \gamma \,(\rho e^{i \lambda})^{-1}  \left\{ \rho e^{i (\mu + \lambda)} 
\right\}^p $.
The $p$th trigonometric moment of a random variable following the wrapped Cauchy distribution is given by $ (\rho_k e^{i \mu_k})^p $ \citep{mcc96}.
Then the following equality holds for the trigonometric moments of the Kato--Jones distribution (\ref{eq:kj_density}) and the wrapped Cauchy mixture
\begin{equation}
\gamma \,(\rho e^{i \lambda})^{-1}  \left\{ \rho e^{i (\mu + \lambda)} \right\}^p = \sum_{k=1}^m \pi_k (\rho_k e^{i \mu_k})^p , \quad p \in \mathbb{N}. \label{eq:tm_wc}
\end{equation}
%This implies that
%\begin{equation}
%\frac{1}{p} \sum_{n=1}^p \gamma (\rho e^{i \lambda})^{-1} \left\{ \rho e^{i (\mu+\lambda-\mu_1)} \right\}^n = \frac{1}{p} \sum_{n=1}^p \sum_{k=1}^m \pi_k \left\{ \rho_k e^{i (\mu_k - \mu_1)} \right\}^n . \label{eq:tm_wc2}
%\end{equation}
Without loss of generality, assume that $\rho_1 \geq \rho_2 \geq \cdots \geq \rho_m$.
%First, we consider the case $\rho >0$.
We will show $\rho=\rho_1$ by contradiction.
First assume $\rho > \rho_1$.
Dividing both sides of the equation (\ref{eq:tm_wc}) by $\{ \rho e^{i (\mu + \lambda )} \}^p$, we have
\begin{equation}
\gamma (\rho e^{i \lambda})^{-1} = \sum_{k=1}^m \pi_k \left\{ \frac{\rho_k}{\rho} e^{i (\mu_k - \mu - \lambda)} \right\}^p.  \label{eq:tm_wc2}
%\gamma (\rho e^{i \lambda})^{-1} \left\{ \frac{\rho}{\rho_1} e^{i (\mu+\lambda-\mu_1)} \right\}^p = \pi_1 + \sum_{k=2}^m \pi_k \left\{ \frac{\rho_k}{\rho_1} e^{i (\mu_k - \mu_1)} \right\}^p. 
%\label{eq:tm_wc2}
\end{equation}
%where
%$
%b_p = \pi_1 + \sum_{k=2}^m \pi_k \{ (\rho_k / \rho_1) e^{i (\mu_k - \mu_1)} \}^p.
%$
Clearly, the left-hand side of (\ref{eq:tm_wc2}) is unchanged to be $\gamma (\rho e^{i \lambda})^{-1}$ as $p \rightarrow \infty$.
On the other hand, since $\rho > \rho_1$,
$$
\lim_{p \rightarrow \infty} \sum_{k=1}^m \pi_k \left\{ \frac{\rho_k}{\rho} e^{i (\mu_k - \mu - \lambda)} \right\}^p = 0.
$$
% Note that, even if there exists $k \geq 2$ such that  $\rho_k = \rho$, $\lim_{p \rightarrow \infty } p^{-1} \sum_{n=1}^p e^{i p (\mu_k-\mu - \lambda)} = 0$ due to the assumption $\mu_k \neq \mu_1$ and Lemma 2.1 of \cite{hol04}.
Thus we have $\gamma=0$, which is contradictory to the assumption $\gamma >0$.
%Therefore $\rho_1 \geq \rho$.
Similarly, if $\rho_1 > \rho$, we can prove the contradiction $\pi_1=0$ by dividing both sides of the equation (\ref{eq:tm_wc}) by $(\rho_1  e^{i \mu_1 })^p$ and taking the limit $p \rightarrow \infty$.
Thus we have $\rho = \rho_1$.

Let us assume that $\rho=\rho_1 = \cdots = \rho_{\ell} > \rho_{\ell+1}$ $(1 \leq \ell \leq m-1)$.
\katov{Since the equation (\ref{eq:tm_wc2}) holds for any $p\geq 1$, we have}
\begin{align}
\gamma (\rho e^{i \lambda})^{-1} & \katov{ = \frac{1}{p} \sum_{n=1}^p \gamma (\rho e^{i \lambda})^{-1} } = \frac{1}{p} \sum_{n=1}^p \sum_{k=1}^m \pi_k \left\{ \frac{\rho_k}{\rho} e^{i (\mu_k - \mu - \lambda)} \right\}^n \label{eq:tm_wc6s} \\
& = \frac{1}{p} \sum_{n=1}^p \sum_{k=1}^{\ell} \pi_k e^{i n (\mu_k - \mu - \lambda)} +  \frac{1}{p} \sum_{n=1}^p  \sum_{k=\ell+1}^m \pi_k \left\{ \frac{\rho_k}{\rho} e^{i (\mu_k - \mu - \lambda)} \right\}^n . \label{eq:tm_wc6}
\end{align}
The second term of (\ref{eq:tm_wc6}) converges to 0 as $p \rightarrow \infty$.
Then
\begin{equation}
\gamma (\rho e^{i \lambda})^{-1} = \lim_{p \rightarrow \infty} \frac{1}{p} \sum_{n=1}^p \sum_{k=1}^{\ell} \pi_k e^{i n (\mu_k - \mu - \lambda)}. \label{eq:tm_wc4}
\end{equation}
It follows from Lemma 2.1 of \cite{hol04} that
\begin{equation}
\lim_{p \rightarrow \infty} \frac{1}{p} \sum_{n=1}^p  e^{i n \theta} = \left\{
\begin{array}{ll}
    1, & \theta=0, \\
    0, & \theta \in [-\pi,\pi) \setminus \{0\}. 
\end{array} \right. \label{eq:hol_lemma}
\end{equation}
This implies that, if $\mu_k - \mu -\lambda \neq 0$ for any $k$, then $\gamma = 0$ which is contradictory to the assumption $\gamma > 0$.
Therefore there exists $\tilde{k} \geq 1$ such that $\mu_{\tilde k} - \mu -\lambda = 0$.
Because $(\mu_k,\rho_k) \neq (\mu_{\ell},\rho_{\ell})$ for $k \neq \ell$, such a $\tilde{k}$ is unique.
Without loss of generality, assume $\tilde{k}=1$.
It follows then that
$$
\gamma (\rho e^{i \lambda})^{-1} = \pi_1.
$$
Since $\pi_1>0$, we have $\pi_1 = \gamma / \rho \in (0,1)$ and $\lambda = 0$.
Substituting this result into the equation (\ref{eq:tm_wc}), we \katov{obtain}
$$
\sum_{k = 2}^m \pi_k \left( \rho_k e^{i \mu_k } \right)^p = 0 . 
$$
This implies that, if $\rho_2 >0$,
\katov{
$$
\pi_2 = - \sum_{k=3}^m \pi_k \left\{ \frac{\rho_k}{\rho_2} e^{i (\mu_k - \mu_2)} \right\}^p.
$$
In a similar manner to how we obtain the equation (\ref{eq:tm_wc6s}) from (\ref{eq:tm_wc2}), we have}
$$
\katov{ \pi_2 = - \frac{1
}{p} \sum_{n=1}^p \sum_{k=3}^m \pi_k \left\{ \frac{\rho_k}{\rho_2} e^{i (\mu_k - \mu_2)} \right\}^n. }
$$
Since \katov{the right-hand side of this equation} converges to \katov{$\pi_2 (\neq 0)$} as $p \rightarrow \infty$, there exists $\tilde{k} \geq 3$ such that $(\mu_{\tilde{k}},\rho_{\tilde{k}})=(\mu_2,\rho_2)$.
However this is the contradiction and we have $\rho_2=0$.
Thus $\rho_2=\cdots= \rho_m=0$.
Summarising the results, the $p$th trigonometric moment of the Kato--Jones distribution (\ref{eq:tm_wc}) has to be of the form
\begin{equation}
\phi_{\rm KJ}(p) = \frac{\gamma}{\rho} (\rho e^{i \mu})^p,  \label{eq:tm_wc5}
\end{equation}
which is equal to the $p$th trigonometric moment of a random variable having the two-component symmetric mixture (\ref{eq:two_wc}).

If $\rho=\rho_1=\cdots = \rho_m$, we obtain the equation (\ref{eq:tm_wc4}) with $\ell =m$ and thus the same trigonometric moment (\ref{eq:tm_wc5}) is derived for the Kato--Jones distribution.
\hfill $\Box$

%As $p \rightarrow \infty$, the left- and right-hand sides of (\ref{eq:tm_wc2}) converge to 0 and $\pi_1 $, respectively.
%Since $\pi_1 > 0$, we have $\rho_1 \geq \rho$.
%Diving both sides of the equation (\ref{eq:tm_wc}) by $(\rho e^{i (\mu + \lambda)})^p$, we have
%$$
%\gamma (\rho e^{i \lambda})^{-1} = \sum_{k=1}^m \pi_k \left\{ \frac{\rho_k}{\rho} e^{i (\mu_k - \mu)} \right\}^p 
%$$
}

%As in the proof of Appendix A, the $p$th trigonometric moment of a random variable $\Theta$ having the Kato--Jones distribution (\ref{eq:kj_density}) is $\phi_{\rm KJ}(p) \equiv E(e^{i p \Theta})$ = \gamma \,(\rho e^{i \lambda})^{-1}  \left\{ \rho e^{i (\mu + \lambda)}  \right\}^p $ \citep{kat15}.

\subsection*{D.3.\quad Proof of Proposition \ref{prop:reparametrization}} \label{sec:proof_reparametrization}
First we show (i).
The parameter space of \chris{the} Kato--Jones distribution (\ref{eq:kj_density}) can be expressed as
$$
-\pi \leq \mu,\lambda < \pi, \quad 0 \leq \rho <1,  \quad 0 \leq \gamma \leq \frac{1+\rho }{2}, \quad \rho \gamma \cos \lambda \geq \frac{\rho^2+2\gamma-1}{2}.
$$
This parameter space is equivalent to
$$
-\pi \leq \mu,\lambda < \pi, \quad 0 \leq \rho <1, \quad 0 \leq \gamma \leq \frac{1-\rho^2}{2(1-\rho \cos \lambda)}.
$$
This expression \chris{for} the parameter space implies that, for given $\rho$ and $\lambda$, $\bar{\gamma} = (1-\rho^2)/\{2(1-\rho \cos \lambda )\} $ achieves the upper bound of the range of $\gamma$.
Using $\overline{\gamma}$, \chris{the} Kato--Jones density (\ref{eq:kj_density}) can be written as
\begin{align*}
	\lefteqn{ g_{\rm KJ}(\theta;\mu,\gamma,\rho,\lambda) } \hspace{0.5cm} \nonumber \\
	& =  \frac{\gamma}{\bar{\gamma}} \cdot \frac{1}{2\pi} \left\{ 1 + 2 \bar{\gamma} \, \frac{\cos 
		(\theta-\mu)  - \rho \cos \lambda}{1+\rho^2-2 \rho \cos 
		(\theta-\mu-\lambda)} \right\} + \left( 1- \frac{\gamma}{\bar{\gamma}} \right) \frac{1}{2\pi} \nonumber \\
		& =  \pi' \cdot \frac{1}{2\pi} \left\{ 1 + 2 \bar{\gamma} \, \frac{\cos 
		(\theta-\mu)  - \rho \cos \lambda}{1+\rho^2-2 \rho \cos 
		(\theta-\mu-\lambda)} \right\} + \left( 1- \pi' \right) \frac{1}{2\pi}, 
		%\label{eq:density2}
\end{align*}
as required.
It is straightforward to show (ii) by substituting expression (\ref{eq:kj_density2}) into the mixture density (\ref{eq:kj_mix_density}). \hfill $\Box$

\katott{
\subsection*{D.4.\quad Proof of Theorem \ref{thm:identifiability}} \label{sec:proof_identifiablity}

Let $\bm{\Psi},\tilde{\bm{\Psi}} \in \Omega$.
Suppose that random variables $\Theta$ and $\tilde{\Theta}$ have the distributions (\ref{eq:kj_mix_density2}) with the parameters $\bm{\Psi}$ and $\tilde{\bm{\Psi}}$, respectively.
We show that if the trigonometric moments of $\Theta$ and $\tilde{\Theta}$ are the same, namely, $E(e^{ip\Theta}) = E(e^{ip \tilde{\Theta}})$ for any $p \in \mathbb{N}$, then $\bm{\Psi} = \tilde{\bm{\Psi}}$.

Write 
\begin{align*}
\bm{\Psi} & = (\mu_1,\ldots,\mu_m,\rho_1,\ldots,\rho_m,\lambda_1,\ldots,\lambda_m,\pi'_1,\ldots,\pi'_{m+1}), \\
\tilde{\bm{\Psi}} &= (\tilde{\mu}_1,\ldots,\tilde{\mu}_{\tilde{m}},\tilde{\rho}_1,\ldots,\tilde{\rho}_{\tilde{m}},\tilde{\lambda}_1,\ldots,\tilde{\lambda}_{\tilde{m}},\tilde{\pi}'_1,\ldots,\tilde{\pi}'_{{\tilde{m}}+1}).
\end{align*}
Then it suffices to show that the following equality holds for any $p \in \mathbb{N}$:
\begin{equation} \label{eq:id_basic}
\sum_{k=1}^m \pi'_k \bar{\gamma}_{k} \,(\rho_k e^{i \lambda_k})^{-1}  \left\{ \rho_k e^{i (\mu_k + \lambda_k )} \right\}^p 
= \sum_{k=1}^{\tilde{m}} \tilde{\pi}'_k \bar{\tilde{\gamma}}_{k} \,(\tilde{\rho}_k e^{i \tilde{\lambda}_k})^{-1}  \left\{ \tilde{\rho}_k e^{i (\tilde{\mu}_k + \tilde{\lambda}_k )} \right\}^p.
\end{equation}

First we prove $\rho_1 = \tilde{\rho}_1$ by contradiction.
Let $\rho_1 > \tilde{\rho}_1$ and divide both sides of equation (\ref{eq:id_basic}) by $\{ \rho_1 e^{i(\mu_1+\lambda_1)} \}^p$.
Then
\begin{align} 
\begin{split} \label{eq:id_rho1}
\lefteqn{ \pi'_1 \bar{\gamma}_{1} \,(\rho_1 e^{i \lambda_1})^{-1} + \sum_{k=2}^m \pi'_k \bar{\gamma}_{k} \,(\rho_k e^{i \lambda_k})^{-1}  \left\{ \frac{\rho_k}{\rho_1} e^{i (\mu_k + \lambda_k - \mu_1 - \lambda_1 )} \right\}^p } \hspace{1cm} \\ 
& = \sum_{k=1}^{\tilde{m}} \tilde{\pi}'_k \bar{\tilde{\gamma}}_{k} \,(\tilde{\rho}_k e^{i \tilde{\lambda}_k})^{-1}  \left\{ \frac{\tilde{\rho}_k}{\rho_1} e^{i (\tilde{\mu}_k + \tilde{\lambda}_k - \mu_1 - \lambda_1 )} \right\}^p.
\end{split}
\end{align}
As $p \rightarrow \infty$, the left- and right-hand sides of (\ref{eq:id_rho1}) converge to $ \pi'_1 \bar{\gamma}_{1} \,(\rho_1 e^{i \lambda_1})^{-1}$ and 0, respectively.
This implies $\pi_1'=0$, which is contradictory to the assumption $\pi_1'>0$.
Similarly, if $\tilde{\rho}_1 > \rho_1 $, then we have the contradiction $\pi_1=0$ by dividing both sides of the equation (\ref{eq:id_basic}) by $\{ \tilde{\rho}_1 e^{i(\tilde{\mu}_1+\tilde{\lambda}_1)} \}^p$ and evaluating the limit $p \rightarrow \infty$.
Therefore we have $\rho_1 = \tilde{\rho}_1$.

Next we show $(\mu_1,\lambda_1,\pi_1')=(\tilde{\mu}_1, \tilde{\lambda}_1, \tilde{\pi}_1')$.
\katov{Applying a similar technique as used in (\ref{eq:tm_wc6s}) to equation (\ref{eq:id_rho1}), it follows that}
\begin{align} 
\begin{split} \label{eq:id_rho1_p}
\lefteqn{ \pi'_1 \bar{\gamma}_{1} \,(\rho_1 e^{i \lambda_1})^{-1} + \frac{1}{p} \sum_{n=1}^p \sum_{k=2}^m \pi'_k \bar{\gamma}_{k} \,(\rho_k e^{i \lambda_k})^{-1}  \left\{ \frac{\rho_k}{\rho_1} e^{i (\mu_k + \lambda_k - \mu_1 - \lambda_1 )} \right\}^n } \hspace{1cm} \\ 
= & \ \frac{1}{p} \sum_{n=1}^p \tilde{\pi}'_1 \bar{\tilde{\gamma}}_1 \,(\rho_1 e^{i \tilde{\lambda}_1})^{-1}  e^{i n (\tilde{\mu}_k + \tilde{\lambda}_k - \mu_1 - \lambda_1 )} \\
& + \frac{1}{p} \sum_{n=1}^p \sum_{k=2}^{\tilde{m}} \tilde{\pi}'_k \bar{\tilde{\gamma}}_{k} \,(\tilde{\rho}_k e^{i \tilde{\lambda}_k})^{-1}  \left\{ \frac{\tilde{\rho}_k}{\rho_1} e^{i (\tilde{\mu}_k + \tilde{\lambda}_k - \mu_1 - \lambda_1 )} \right\}^n.
\end{split}
\end{align}
The second terms of both sides of (\ref{eq:id_rho1_p}) converge to 0 as $p \rightarrow \infty$.
It follows that
$$
 \lim_{p \rightarrow \infty } \frac{1}{p} \sum_{n=1}^p \tilde{\pi}'_1 \bar{\tilde{\gamma}}_1 \,(\rho_1 e^{i \tilde{\lambda}_1})^{-1}  e^{i n (\tilde{\mu}_k + \tilde{\lambda}_k - \mu_1 - \lambda_1 )} =  \pi'_1 \bar{\gamma}_{1} \,(\rho_1 e^{i \lambda_1})^{-1}.
$$
The equation (\ref{eq:hol_lemma}) implies that if $\tilde{\mu}_1+\tilde{\lambda}_1 \neq \mu_1 + \lambda_1$, then we have $\pi_1'=0$ which is a contradiction.
Thus $\tilde{\mu}_1+\tilde{\lambda}_1 = \mu_1 + \lambda_1$.
Thus
$$
\pi'_1 \bar{\gamma}_{1} \,(\rho_1 e^{i \lambda_1})^{-1} = \tilde{\pi}'_1 \bar{\tilde{\gamma}}_1 \,(\rho_1 e^{i \tilde{\lambda}_1})^{-1}.
$$
Remembering $\bar{\gamma}_1 = (1-\rho_1^2)/\{2(1-\rho_1 \cos \lambda_1 )\} $, we have $(\mu_1,\lambda_1,\pi_1')=(\tilde{\mu}_1, \tilde{\lambda}_1, \tilde{\pi}_1')$.

Since all the parameters and mixing proportion of the first components of both mixtures are the same, the equation (\ref{eq:id_basic}) reduces to
\begin{equation} \label{eq:id_basic2}
\sum_{k=2}^m \pi'_k \bar{\gamma}_{k} \,(\rho_k e^{i \lambda_k})^{-1}  \left\{ \rho_k e^{i (\mu_k + \lambda_k )} \right\}^p 
= \sum_{k=2}^{\tilde{m}} \tilde{\pi}'_k \bar{\tilde{\gamma}}_{k} \,(\tilde{\rho}_k e^{i \tilde{\lambda}_k})^{-1}  \left\{ \tilde{\rho}_k e^{i (\tilde{\mu}_k + \tilde{\lambda}_k )} \right\}^p.
\end{equation}
From this expression, it is straightforward to see $(\mu_k,\rho_k,\lambda_k,\pi_k') = (\tilde{\mu}_k, \tilde{\rho}_k, \tilde{\lambda}_k, \tilde{\pi}_k')$ $(k=2,\ldots,\min \{m , \tilde{m}\} )$ by induction.

Finally, we show $m=\tilde{m}$.
If $m > \tilde{m}$, the equation (\ref{eq:id_basic2}) reduces to
$$ 
\sum_{k=\tilde{m}+1}^m \pi'_k \bar{\gamma}_{k} \,(\rho_k e^{i \lambda_k})^{-1}  \left\{ \rho_k e^{i (\mu_k + \lambda_k )} \right\}^p 
= 0.
$$
If $m=\tilde{m}+1$, it is straightforward to see $\pi_m'=0$.  
If $m > \tilde{m}+1$, then
\begin{equation} \label{eq:id_last}
\pi'_{\tilde{m}+1} \bar{\gamma}_{\tilde{m}+1} \,(\rho_{\tilde{m}+1} e^{i \lambda_{\tilde{m}+1}})^{-1} = - \sum_{k=\tilde{m}+2}^m \pi'_k \bar{\gamma}_{k} \,(\rho_k e^{i \lambda_k})^{-1}  \left\{ \frac{\rho_k}{\rho_{\tilde{m}+1}} e^{i (\mu_k + \lambda_k - \mu_{\tilde{m}+1} - \lambda_{\tilde{m}+1} )} \right\}^p.
\end{equation}
Since the right-hand side of (\ref{eq:id_last}) converges to $0$ as $p \rightarrow \infty$, we have $\pi'_{\tilde{m}+1}=0$.
In a similar manner, for $\tilde{m} < m $, we can show the contradiction $\tilde{\pi}'_{\tilde{m}} = 0$ or $\tilde{\pi}'_{m+1}=0$.
Therefore $m=\tilde{m}$.

Summarising the results, we obtain $\bm{\Psi}=\tilde{\bm{\Psi}}$ as required. \hfill $\Box$
}